\input harvmac.tex
 \input epsf.tex
 \input amssym
 \input color
 \input

\def\figin{\epsfcheck\figin}\def\figins{\epsfcheck\figins}
\def\epsfcheck{\ifx\epsfbox\UnDeFiNeD
\message{(NO epsf.tex, FIGURES WILL BE IGNORED)}
\gdef\figin##1{\vskip2in}\gdef\figins##1{\hskip.5in}
\else\message{(FIGURES WILL BE INCLUDED)}%
\gdef\figin##1{##1}\gdef\figins##1{##1}\fi}
\def\DefWarn#1{}
\def\figinsert{\goodbreak\midinsert}
\def\ifig#1#2#3{\DefWarn#1\xdef#1{fig.~\the\figno}
\writedef{#1\leftbracket fig.\noexpand~\the\figno} %
\figinsert\figin{\centerline{#3}}\medskip\centerline{\vbox{\baselineskip12pt
\advance\hsize by -1truein\noindent\footnotefont{\bf
Fig.~\the\figno:} #2}}
\bigskip\endinsert\global\advance\figno by1}


\def \pa {\partial}
\def \d {\partial}

\def \eps {\epsilon}
\def \( {\left(}
\def \) {\right)}
\def \[ {\left[}
\def \] {\right]}

\lref\AharonyIPA{
  O.~Aharony and Z.~Komargodski,
  ``The Effective Theory of Long Strings,''
JHEP {\bf 1305}, 118 (2013).
[arXiv:1302.6257 [hep-th]].
}

\lref\SimmonsDuffinGJK{
  D.~Simmons-Duffin,
  ``The Conformal Bootstrap,''
[arXiv:1602.07982 [hep-th]].
}

\lref\RychkovIQZ{
  S.~Rychkov,
  ``EPFL Lectures on Conformal Field Theory in D>= 3 Dimensions,''
[arXiv:1601.05000 [hep-th]].
}

\lref\VenezianoYB{
  G.~Veneziano,
  ``Construction of a crossing - symmetric, Regge behaved amplitude for linearly rising trajectories,''
Nuovo Cim.\ A {\bf 57}, 190 (1968).
}

\lref\FitzpatrickYX{
  A.~L.~Fitzpatrick, J.~Kaplan, D.~Poland and D.~Simmons-Duffin,
``The Analytic Bootstrap and AdS Superhorizon Locality,''
JHEP {\bf 1312}, 004 (2013).
[arXiv:1212.3616 [hep-th]].
}

\lref\KomargodskiEK{
  Z.~Komargodski and A.~Zhiboedov,
  ``Convexity and Liberation at Large Spin,''
JHEP {\bf 1311}, 140 (2013).
[arXiv:1212.4103 [hep-th]].
}

\lref\KomargodskiGCI{
  Z.~Komargodski, M.~Kulaxizi, A.~Parnachev and A.~Zhiboedov,
  ``Conformal Field Theories and Deep Inelastic Scattering,''
[arXiv:1601.05453 [hep-th]].
}

\lref\HellermanCBA{
  S.~Hellerman, S.~Maeda, J.~Maltz and I.~Swanson,
  ``Effective String Theory Simplified,''
JHEP {\bf 1409}, 183 (2014).
[arXiv:1405.6197 [hep-th]].
}

\lref\SiversIG{
  D.~Sivers and J.~Yellin,
  ``Review of recent work on narrow resonance models,''
Rev.\ Mod.\ Phys.\  {\bf 43}, 125 (1971).
}

\lref\PandoZayasYB{
  L.~A.~Pando Zayas, J.~Sonnenschein and D.~Vaman,
  ``Regge trajectories revisited in the gauge / string correspondence,''
Nucl.\ Phys.\ B {\bf 682}, 3 (2004).
[hep-th/0311190].
}

\lref\KarchPV{
  A.~Karch, E.~Katz, D.~T.~Son and M.~A.~Stephanov,
  ``Linear confinement and AdS/QCD,''
Phys.\ Rev.\ D {\bf 74}, 015005 (2006).
[hep-ph/0602229].
}

\lref\RychkovIQZ{
  S.~Rychkov,
  ``EPFL Lectures on Conformal Field Theory in D>= 3 Dimensions,''
[arXiv:1601.05000 [hep-th]].
}

\lref\BrowerEA{
  R.~C.~Brower, J.~Polchinski, M.~J.~Strassler and C.~I.~Tan,
  ``The Pomeron and gauge/string duality,''
JHEP {\bf 0712}, 005 (2007).
[hep-th/0603115].
}

\lref\mandelstam{
S.~Mandelstam, ``Dual-resonance models." Physics Reports 13.6 (1974): 259-353.
}

\lref\FreundHW{
  P.~G.~O.~Freund,
  ``Finite energy sum rules and bootstraps,''
Phys.\ Rev.\ Lett.\  {\bf 20}, 235 (1968).
}

\lref\MeyerJC{
  H.~B.~Meyer and M.~J.~Teper,
  ``Glueball Regge trajectories and the pomeron: A Lattice study,''
Phys.\ Lett.\ B {\bf 605}, 344 (2005).
[hep-ph/0409183].
}

\lref\CoonYW{
  D.~D.~Coon,
  ``Uniqueness of the veneziano representation,''
Phys.\ Lett.\ B {\bf 29}, 669 (1969).
}

\lref\FairlieAD{
  D.~B.~Fairlie and J.~Nuyts,
  ``A fresh look at generalized Veneziano amplitudes,''
Nucl.\ Phys.\ B {\bf 433}, 26 (1995).
[hep-th/9406043].
}

\lref\PonomarevJQK{
  D.~Ponomarev and A.~A.~Tseytlin,
  ``On quantum corrections in higher-spin theory in flat space,''
[arXiv:1603.06273 [hep-th]].
}

\lref\StromingerTalk{
  A.~Strominger, Talk at Strings 2014, Princeton.
}

\lref\CostaMG{
  M.~S.~Costa, J.~Penedones, D.~Poland and S.~Rychkov,
  ``Spinning Conformal Correlators,''
JHEP {\bf 1111}, 071 (2011).
[arXiv:1107.3554 [hep-th]].
}

\lref\CamanhoAPA{
  X.~O.~Camanho, J.~D.~Edelstein, J.~Maldacena and A.~Zhiboedov,
  ``Causality Constraints on Corrections to the Graviton Three-Point Coupling,''
JHEP {\bf 1602}, 020 (2016).
[arXiv:1407.5597 [hep-th]].
}

\lref\GrossKZA{
  D.~J.~Gross and P.~F.~Mende,
  ``The High-Energy Behavior of String Scattering Amplitudes,''
Phys.\ Lett.\ B {\bf 197}, 129 (1987).
}

\lref\KarlinerHD{
  M.~Karliner, I.~R.~Klebanov and L.~Susskind,
  ``Size and Shape of Strings,''
Int.\ J.\ Mod.\ Phys.\ A {\bf 3}, 1981 (1988).
}

\lref\SusskindAA{
  L.~Susskind,
  ``Strings, black holes and Lorentz contraction,''
Phys.\ Rev.\ D {\bf 49}, 6606 (1994).
[hep-th/9308139].
}

\lref\tHooftJZ{
  G.~'t Hooft,
  ``A Planar Diagram Theory for Strong Interactions,''
Nucl.\ Phys.\ B {\bf 72}, 461 (1974).
}

\lref\WittenKH{
  E.~Witten,
  ``Baryons in the 1/n Expansion,''
Nucl.\ Phys.\ B {\bf 160}, 57 (1979).
}

\lref\zeros{
I.~E.~Pritsker and A.~M.~Yeager,
``Zeros of Polynomials with Random Coefficients,"
J. Approx. Theory 189 (2015), 88-100.}

\lref\zerosort{
W.~V.~Assche,
``Some results on the asymptotic distribution of the zeros of orthogonal polynomials,"
Comput. Math. Appl. 12-13 (1985) 615-623.
}

\lref\Brower{
R.~C.~Brower and J.~Harte. 
``Kinematic Constraints for Infinitely Rising Regge Trajectories," 
Physical Review 164.5 (1967): 1841.
}

\lref\PolchinskiUF{
  J.~Polchinski and M.~J.~Strassler,
  ``The String dual of a confining four-dimensional gauge theory,''
[hep-th/0003136].
}

\lref\PappadopuloJK{
  D.~Pappadopulo, S.~Rychkov, J.~Espin and R.~Rattazzi,
  ``OPE Convergence in Conformal Field Theory,''
Phys.\ Rev.\ D {\bf 86}, 105043 (2012).
[arXiv:1208.6449 [hep-th]].
}

\lref\GiddingsGJ{
  S.~B.~Giddings and R.~A.~Porto,
  ``The Gravitational S-matrix,''
Phys.\ Rev.\ D {\bf 81}, 025002 (2010).
[arXiv:0908.0004 [hep-th]].
}

\lref\HellermanKBA{
  S.~Hellerman and I.~Swanson,
  ``String Theory of the Regge Intercept,''
Phys.\ Rev.\ Lett.\  {\bf 114}, no. 11, 111601 (2015).
[arXiv:1312.0999 [hep-th]].
}

\lref\AharonyIPA{
  O.~Aharony and Z.~Komargodski,
  ``The Effective Theory of Long Strings,''
JHEP {\bf 1305}, 118 (2013).
[arXiv:1302.6257 [hep-th]].
}

\lref\PolchinskiAX{
  J.~Polchinski and A.~Strominger,
  ``Effective string theory,''
Phys.\ Rev.\ Lett.\  {\bf 67}, 1681 (1991).
}

\lref\JainNZA{
  S.~Jain, M.~Mandlik, S.~Minwalla, T.~Takimi, S.~R.~Wadia and S.~Yokoyama,
 ``Unitarity, Crossing Symmetry and Duality of the S-matrix in large N Chern-Simons theories with fundamental matter,''
JHEP {\bf 1504}, 129 (2015).
[arXiv:1404.6373 [hep-th]].
}

\lref\DubovskySH{
  S.~Dubovsky, R.~Flauger and V.~Gorbenko,
  ``Effective String Theory Revisited,''
JHEP {\bf 1209}, 044 (2012).
[arXiv:1203.1054 [hep-th]].
}

\lref\LegRed{
Askey, Richard. "Orthogonal expansions with positive coefficients." Proceedings of the American Mathematical Society 16.6 (1965): 1191-1194.
}

\lref\AmatiWQ{
  D.~Amati, M.~Ciafaloni and G.~Veneziano,
  ``Superstring Collisions at Planckian Energies,''
Phys.\ Lett.\ B {\bf 197}, 81 (1987).
}

\lref\MatsudaKF{
  S.~Matsuda,
  ``Uniqueness of the veneziano representation,''
Phys.\ Rev.\  {\bf 185}, 1811 (1969).
}

\lref\KhuriAX{
  N.~N.~Khuri,
  ``Derivation of a veneziano series from the regge representation,''
Phys.\ Rev.\  {\bf 185}, 1876 (1969).
}

\lref\WeimarPI{
  E.~Weimar,
  ``Alternatives to the Veneziano Amplitude,''
    DESY-74-3.
}

\lref\WandersET{
  G.~Wanders,
  ``Constraints on the zeros and the asymptotic behavior of a veneziano amplitude,''
Phys.\ Lett.\ B {\bf 34}, 325 (1971).
}

\lref\GrossGE{
  D.~J.~Gross and J.~L.~Manes,
``The High-energy Behavior of Open String Scattering,''
Nucl.\ Phys.\ B {\bf 326}, 73 (1989).
}

\lref\ChodosGT{
  A.~Chodos and C.~B.~Thorn,
  ``Making the Massless String Massive,''
Nucl.\ Phys.\ B {\bf 72}, 509 (1974).
}

\lref\SonnenscheinPIM{
  J.~Sonnenschein,
  ``Holography Inspired Stringy Hadrons,''
[arXiv:1602.00704 [hep-th]].
}

\lref\ElShowkHT{
  S.~El-Showk, M.~F.~Paulos, D.~Poland, S.~Rychkov, D.~Simmons-Duffin and A.~Vichi,
  ``Solving the 3D Ising Model with the Conformal Bootstrap,''
Phys.\ Rev.\ D {\bf 86}, 025022 (2012).
[arXiv:1203.6064 [hep-th]].
}

\lref\KosBKA{
  F.~Kos, D.~Poland and D.~Simmons-Duffin,
  ``Bootstrapping Mixed Correlators in the 3D Ising Model,''
JHEP {\bf 1411}, 109 (2014).
[arXiv:1406.4858 [hep-th]].
}

\lref\KlebanovJA{
  I.~R.~Klebanov and A.~M.~Polyakov,
``AdS dual of the critical O(N) vector model,''
Phys.\ Lett.\ B {\bf 550}, 213 (2002).
[hep-th/0210114].
}

\lref\KorchemskyRC{
  G.~P.~Korchemsky, J.~Kotanski and A.~N.~Manashov,
``Multi-reggeon compound states and resummed anomalous dimensions in QCD,''
  Phys.\ Lett.\ B {\bf 583} (2004) 121
  doi:10.1016/j.physletb.2004.01.014
  [hep-ph/0306250].
}

\lref\NimaYutin{
N.~Arkani-Hamed, Y.~Huang and T.~C.~Huang,
``String theory as the unique weakly coupled UV Completion of YM and GR," In progress}

\lref\JoaoPedro{
M.~ Paulos, J.~Penedones, J.~Toledo, B.~C.~van Rees and P.~Vieira 
``2d S-matrix Bootstrap", To appear
}

\lref\Haus{F.~Hausdorff, ``Momentprobleme f\"ur ein endliches Intervall.'' Mathematische Zeitschrift 16.1 (1923): 220-248.}

\lref\PenedonesUE{
  J.~Penedones,
  ``Writing CFT correlation functions as AdS scattering amplitudes,''
JHEP {\bf 1103}, 025 (2011).
[arXiv:1011.1485 [hep-th]].
}

\lref\VasilievBA{
  M.~A.~Vasiliev,
``Higher spin gauge theories: Star product and AdS space,''
In *Shifman, M.A. (ed.): The many faces of the superworld* 533-610.
[hep-th/9910096].
}
\lref\MaldacenaSF{
  J.~Maldacena and A.~Zhiboedov,
  ``Constraining conformal field theories with a slightly broken higher spin symmetry,''
Class.\ Quant.\ Grav.\  {\bf 30}, 104003 (2013).
[arXiv:1204.3882 [hep-th]].
}

\lref\Tao{
T.~Tao, V.~Vu. 
``Local universality of zeroes of random polynomials." 
International Mathematics Research Notices (2014): rnu084,
[arXiv:1307.4357 [math.PR]].
}

\lref\MakeenkoRF{
  Y.~Makeenko and P.~Olesen,
  ``Wilson Loops and QCD/String Scattering Amplitudes,''
Phys.\ Rev.\ D {\bf 80}, 026002 (2009).
[arXiv:0903.4114 [hep-th]].
}

\lref\ArmoniNJA{
  A.~Armoni,
  ``Large-N QCD and the Veneziano Amplitude,''
Phys.\ Lett.\ B {\bf 756}, 328 (2016).
[arXiv:1509.03077 [hep-th]].
}

\lref\ArkaniHamedBZA{
  N.~Arkani-Hamed and J.~Maldacena,
  ``Cosmological Collider Physics,''
[arXiv:1503.08043 [hep-th]].
}

\lref\HeemskerkPN{
  I.~Heemskerk, J.~Penedones, J.~Polchinski and J.~Sully,
  ``Holography from Conformal Field Theory,''
JHEP {\bf 0910}, 079 (2009).
[arXiv:0907.0151 [hep-th]].
}

\lref\PenedonesUE{
  J.~Penedones,
  ``Writing CFT correlation functions as AdS scattering amplitudes,''
JHEP {\bf 1103}, 025 (2011).
[arXiv:1011.1485 [hep-th]].
}

\lref\AldayZY{
  L.~F.~Alday, B.~Eden, G.~P.~Korchemsky, J.~Maldacena and E.~Sokatchev,
JHEP {\bf 1109}, 123 (2011).
[arXiv:1007.3243 [hep-th]].
}

\lref\CostaCB{
  M.~S.~Costa, V.~Goncalves and J.~Penedones,
  ``Conformal Regge theory,''
JHEP {\bf 1212}, 091 (2012).
[arXiv:1209.4355 [hep-th]].
}

\lref\MaldacenaWAA{
  J.~Maldacena, S.~H.~Shenker and D.~Stanford,
  ``A bound on chaos,''
JHEP {\bf 1608}, 106 (2016).
[arXiv:1503.01409 [hep-th]].
}

\lref\CaronHuotICG{
  S.~Caron-Huot, Z.~Komargodski, A.~Sever and A.~Zhiboedov,
 ``The Asymptotic Uniqueness of the Veneziano Amplitude,'' 
[arXiv:1607.04253 [hep-th]].
}

\lref\AldayMF{
  L.~F.~Alday and J.~M.~Maldacena,
  ``Comments on operators with large spin,''
JHEP {\bf 0711}, 019 (2007).
[arXiv:0708.0672 [hep-th]].
}

\lref\AldayZY{
  L.~F.~Alday, B.~Eden, G.~P.~Korchemsky, J.~Maldacena and E.~Sokatchev,
  ``From correlation functions to Wilson loops,''
JHEP {\bf 1109}, 123 (2011).
[arXiv:1007.3243 [hep-th]].
}

\lref\HartmanLFA{
  T.~Hartman, S.~Jain and S.~Kundu,
  ``Causality Constraints in Conformal Field Theory,''
JHEP {\bf 1605}, 099 (2016).
[arXiv:1509.00014 [hep-th]].
}

\lref\MackMI{
  G.~Mack,
  ``D-independent representation of Conformal Field Theories in D dimensions via transformation to auxiliary Dual Resonance Models. Scalar amplitudes,''
[arXiv:0907.2407 [hep-th]].
}

\lref\KomargodskiEK{
  Z.~Komargodski and A.~Zhiboedov,
  ``Convexity and Liberation at Large Spin,''
JHEP {\bf 1311}, 140 (2013).
[arXiv:1212.4103 [hep-th]].
}

\lref\FitzpatrickYX{
  A.~L.~Fitzpatrick, J.~Kaplan, D.~Poland and D.~Simmons-Duffin,
  ``The Analytic Bootstrap and AdS Superhorizon Locality,''
JHEP {\bf 1312}, 004 (2013).
[arXiv:1212.3616 [hep-th]].
}

\lref\GopakumarWKT{
  R.~Gopakumar, A.~Kaviraj, K.~Sen and A.~Sinha,
  ``Conformal Bootstrap in Mellin Space,''
[arXiv:1609.00572 [hep-th]].
}

\lref\AldayHR{
  L.~F.~Alday and J.~M.~Maldacena,
  ``Gluon scattering amplitudes at strong coupling,''
JHEP {\bf 0706}, 064 (2007).
[arXiv:0705.0303 [hep-th]].
}

\lref\PolchinskiAX{
  J.~Polchinski and A.~Strominger,
  ``Effective string theory,''
Phys.\ Rev.\ Lett.\  {\bf 67}, 1681 (1991).
}

\lref\HellermanKBA{
  S.~Hellerman and I.~Swanson,
  ``String Theory of the Regge Intercept,''
Phys.\ Rev.\ Lett.\  {\bf 114}, no. 11, 111601 (2015).
[arXiv:1312.0999 [hep-th]].
}

\lref\AharonyDB{
  O.~Aharony and N.~Klinghoffer,
  ``Corrections to Nambu-Goto energy levels from the effective string action,''
JHEP {\bf 1012}, 058 (2010).
[arXiv:1008.2648 [hep-th]].
}

\lref\BarbashovNQ{
  B.~M.~Barbashov and V.~V.~Nesterenko,
  ``Relativistic String with Massive Ends,''
Theor.\ Math.\ Phys.\  {\bf 31}, 465 (1977), [Teor.\ Mat.\ Fiz.\  {\bf 31}, 291 (1977)].
}

\lref\mandelstam{
S.~Mandelstam, ``Dual-resonance models." Physics Reports 13.6 (1974): 259-353.
}

\lref\VasilievEN{
  M.~A.~Vasiliev,
  ``Consistent equation for interacting gauge fields of all spins in (3+1)-dimensions,''
Phys.\ Lett.\ B {\bf 243}, 378 (1990).
}

\lref\MooreNS{
  G.~W.~Moore,
  ``Symmetries of the bosonic string S matrix,''
[hep-th/9404025, hep-th/9310026].
}

\lref\CostaCB{
  M.~S.~Costa, V.~Goncalves and J.~Penedones,
  ``Conformal Regge theory,''
JHEP {\bf 1212}, 091 (2012).
[arXiv:1209.4355 [hep-th]].
}

\lref\VenezianoYB{
  G.~Veneziano,
  ``Construction of a crossing - symmetric, Regge behaved amplitude for linearly rising trajectories,''
Nuovo Cim.\ A {\bf 57}, 190 (1968).
}

\lref\AndreevSY{
  O.~Andreev and W.~Siegel,
  ``Quantized tension: Stringy amplitudes with Regge poles and parton behavior,''
Phys.\ Rev.\ D {\bf 71}, 086001 (2005).
[hep-th/0410131].
}

\lref\VenezianoCKS{
  G.~Veneziano, S.~Yankielowicz and E.~Onofri,
  ``A model for pion-pion scattering in large-N QCD,''
JHEP {\bf 1704}, 151 (2017).
[arXiv:1701.06315 [hep-th]].
}

\lref\VirasoroME{
  M.~A.~Virasoro,
  ``Alternative constructions of crossing-symmetric amplitudes with regge behavior,''
Phys.\ Rev.\  {\bf 177}, 2309 (1969).
}

\lref\ShapiroGY{
  J.~A.~Shapiro,
  ``Electrostatic analog for the virasoro model,''
Phys.\ Lett.\  {\bf 33B}, 361 (1970).
}

\lref\PaulosFAP{
  M.~F.~Paulos, J.~Penedones, J.~Toledo, B.~C.~van Rees and P.~Vieira,
  ``The S-matrix Bootstrap I: QFT in AdS,''
[arXiv:1607.06109 [hep-th]].
}

\lref\PaulosBUT{
  M.~F.~Paulos, J.~Penedones, J.~Toledo, B.~C.~van Rees and P.~Vieira,
  ``The S-matrix Bootstrap II: Two Dimensional Amplitudes,''
[arXiv:1607.06110 [hep-th]].
}

\lref\FubiniQB{
  S.~Fubini and G.~Veneziano,
  ``Level structure of dual-resonance models,''
Nuovo Cim.\ A {\bf 64}, 811 (1969).
}

\lref\FubiniWP{
  S.~Fubini, D.~Gordon and G.~Veneziano,
  ``A general treatment of factorization in dual resonance models,''
Phys.\ Lett.\  {\bf 29B}, 679 (1969).
}

\lref\PolchinskiTT{
  J.~Polchinski and M.~J.~Strassler,
  ``Hard scattering and gauge / string duality,''
Phys.\ Rev.\ Lett.\  {\bf 88}, 031601 (2002).
[hep-th/0109174].
}

\lref\ErdmengerCM{
  J.~Erdmenger, N.~Evans, I.~Kirsch and E.~Threlfall,
  ``Mesons in Gauge/Gravity Duals - A Review,''
Eur.\ Phys.\ J.\ A {\bf 35}, 81 (2008).
[arXiv:0711.4467 [hep-th]].
}

\lref\KruczenskiME{
  M.~Kruczenski, L.~A.~Pando Zayas, J.~Sonnenschein and D.~Vaman,
  ``Regge trajectories for mesons in the holographic dual of large-N(c) QCD,''
JHEP {\bf 0506}, 046 (2005).
[hep-th/0410035].
}

\lref\PaulosFAP{
  M.~F.~Paulos, J.~Penedones, J.~Toledo, B.~C.~van Rees and P.~Vieira,
  ``The S-matrix Bootstrap I: QFT in AdS,''
[arXiv:1607.06109 [hep-th]].
}

\lref\PaulosBUT{
  M.~F.~Paulos, J.~Penedones, J.~Toledo, B.~C.~van Rees and P.~Vieira,
  ``The S-matrix Bootstrap II: Two Dimensional Amplitudes,''
[arXiv:1607.06110 [hep-th]].
}

\lref\GiombiMS{
  S.~Giombi and X.~Yin,
  ``The Higher Spin/Vector Model Duality,''
J.\ Phys.\ A {\bf 46}, 214003 (2013).
[arXiv:1208.4036 [hep-th]].
}

\lref\MuruganETO{
  J.~Murugan, D.~Stanford and E.~Witten,
  ``More on Supersymmetric and 2d Analogs of the SYK Model,''
[arXiv:1706.05362 [hep-th]].
}

\lref\DubovskyGI{
  S.~Dubovsky, R.~Flauger and V.~Gorbenko,
  ``Evidence from Lattice Data for a New Particle on the Worldsheet of the QCD Flux Tube,''
Phys.\ Rev.\ Lett.\  {\bf 111}, no. 6, 062006 (2013).
[arXiv:1301.2325 [hep-th]].
}

\lref\DubovskyZEY{
  S.~Dubovsky and V.~Gorbenko,
  ``Towards a Theory of the QCD String,''
JHEP {\bf 1602}, 022 (2016).
[arXiv:1511.01908 [hep-th]].
}

\lref\SelemND{
  A.~Selem and F.~Wilczek,
  ``Hadron systematics and emergent diquarks,''
[hep-ph/0602128].
}

\lref\SonnenscheinJWA{
  J.~Sonnenschein and D.~Weissman,
  ``Rotating strings confronting PDG mesons,''
JHEP {\bf 1408}, 013 (2014).
[arXiv:1402.5603 [hep-ph]].
}

\lref\GopakumarCPB{
  R.~Gopakumar, A.~Kaviraj, K.~Sen and A.~Sinha,
  ``A Mellin space approach to the conformal bootstrap,''
JHEP {\bf 1705}, 027 (2017).
[arXiv:1611.08407 [hep-th]].
}

\lref\PolchinskiTT{
  J.~Polchinski and M.~J.~Strassler,
  ``Hard scattering and gauge / string duality,''
Phys.\ Rev.\ Lett.\  {\bf 88}, 031601 (2002).
[hep-th/0109174].
}

\lref\ElShowkHT{
  S.~El-Showk, M.~F.~Paulos, D.~Poland, S.~Rychkov, D.~Simmons-Duffin and A.~Vichi,
  ``Solving the 3D Ising Model with the Conformal Bootstrap,''
Phys.\ Rev.\ D {\bf 86}, 025022 (2012).
[arXiv:1203.6064 [hep-th]].
}

\lref\GrossUE{
  D.~J.~Gross,
  ``High-Energy Symmetries of String Theory,''
Phys.\ Rev.\ Lett.\  {\bf 60}, 1229 (1988).
}

\lref\PolyakovCS{
  A.~M.~Polyakov,
  ``Fine Structure of Strings,''
Nucl.\ Phys.\ B {\bf 268}, 406 (1986).
}

\Title{
\vbox{\baselineskip6pt
}}
{\vbox{
\centerline{On Fine Structure of Strings:}
\vskip .2in
\centerline{The Universal Correction to the Veneziano Amplitude}
}}

\bigskip
\centerline{Amit Sever$^{1,2}$ and Alexander Zhiboedov$^{3}$}
\bigskip
\centerline{\it $^{1}$ CERN, Theoretical Physics Department, 1211 Geneva 23, Switzerland }
\centerline{\it $^{2}$ School of Physics and Astronomy, Tel Aviv University, Ramat Aviv 69978, Israel}
\centerline{\it $^{3}$ Department of Physics, Harvard University, Cambridge, MA 20138, USA
}

\vskip .2in 

\noindent
We consider theories of weakly interacting higher spin particles in flat spacetime. We focus on the four-point scattering amplitude at high energies and imaginary scattering angles. The leading asymptotic of the amplitude in this regime is universal and equal to the corresponding limit of the Veneziano amplitude. In this paper, we find that the first sub-leading correction to this asymptotic is universal as well. We compute the correction using a model of relativistic strings with massive endpoints. We argue that it is unique using holography, effective theory of long strings and bootstrap techniques.
\Date{ }

\listtoc\writetoc
\vskip .9in \noindent


\newsec{Introduction}

Weakly interacting higher spin particles (WIHS) appear in very different physical contexts \PolyakovCS. One is the Yang-Mills theory at large $N$, where higher spin particles correspond to approximately stable resonances (glueballs) with various masses and spins. Another is tree-level scattering amplitudes of fundamental strings. In this case, higher spin particles describe excitations of a fundamental string.

The $S$-matrix bootstrap aims to classify all such theories of weakly interacting higher spin particles (WIHS) based on general principles. Two-to-two scattering in such theories is described by a meromorphic function $A(s,t)$. It packs together in a neat way the spectrum of particles $m_i$ and a set of three-point couplings $c_{i j k}$.

Unitarity and causality, together with the presence of particles of spin higher than two, results in a set of highly non-trivial constraints on the amplitude. At the present moment only one solution to the problem is known explicitly -- the celebrated Veneziano formula \VenezianoYB\ (and its cousins \refs{\VirasoroME,\ShapiroGY}). This amplitude has several non-generic properties such as exact linearity of the Regge trajectories, an exact degeneracy of the spectrum and soft high energy, fixed angle behavior \GrossKZA.\foot{ In \refs{\AndreevSY,\VenezianoCKS} an ansatz for the scattering amplitude was analyzed for which the leading Regge trajectory is piece-wise linear and flat for negative $t$. It consists of an infinite sum of the Veneziano amplitudes with different slopes approaching zero, all sharing the same mass spectrum.} Without further assumptions, (i.e. the existence of massless spin two particle in the spectrum), these properties are not expected to hold. In other known theories of WIHS, such as confining gauge theories in the large $N$ limit, the high energy behavior for real scattering angles is power-like \PolchinskiTT\ and no degeneracy at all is expected  
in the spectrum \refs{\MeyerJC,\DubovskySH,\AharonyDB}. Correspondingly, the details of $A(s,t)$ may differ significantly from one WIHS to another.

In \CaronHuotICG\ a systematic study of WIHS theories was initiated by studying the scattering amplitudes at large energies and imaginary scattering angles. This kinematical regime is controlled by the fastest spinning particles and was shown to be universal \CaronHuotICG. Namely, in any WIHS the $s,t \gg 1$ asymptotic of the amplitude $A(s,t)$ coincides with the limit of the Veneziano amplitude
\eqn\limitVen{
\lim_{s,t \gg 1}\log A(s,t) = \alpha' \left[ (s+t) \log (s+t) - s \log s - t \log t \right] + ... \ .
}

The analysis of \CaronHuotICG\ assumed that there is no accumulation point in the spectrum and that there is a separation of scales in the large $s,t$ limit. Correspondingly, the result \limitVen\ shows that any WIHS theory satisfying these assumptions is a theory of strings. Specifically, its spectrum contains an infinite set of asymptotically linear Regge trajectories. 

To put this bootstrap program on a systematic computational path, we must understand what the possible corrections to \limitVen\ are. Universality can only arise for the first few terms in the large $s,t$ expansion. Beyond the universal terms, more assumptions will be needed to fix the amplitude and the corresponding class of WIHS theories. 

The aim of this paper is to study the universal corrections to \limitVen. The result of our analysis is that there is a unique, universal correction to \limitVen\ that grows with energy. This correction scales as $s^{1/4}$ in the limit $s,t \gg 1$ with ${s \over t}$ held fixed. It takes the following form\foot{A formula for the $n$-point function could be found in the main text.}
\eqn\mainresult{
\delta \log A(s,t) = -m^{3/2} {16 \sqrt{\pi} \alpha' \over 3} \left( {s t \over s + t} \right)^{1/4} \left[ K\left({s \over s + t} \right) + K \left({t \over s + t} \right) \right] + ... \ ,
}
where $K(x)$ is a complete elliptic integral of the first kind. A parameter $m$ is the new scale that enters the amplitude. On the string theory side, it could be thought of as adding a point mass at the endpoint of the string that slows down its motion.\foot{For application of this model to the experimental data see \SonnenscheinJWA.} On the bootstrap side it is associated with the way the degeneracy of the spectrum is lifted. The correction \mainresult\ implies that the leading Regge trajectory takes the following form
\eqn\reggelead{
j(t) = \alpha'  \left(t -  {8 \sqrt{\pi} \over 3} m^{3/2} t^{1/4} + O(1) \right) \ .
}
The same statement is true about an infinite set of subleading Regge trajectories out of which \limitVen\ and \mainresult\ are built. 

The paper is organized as follows. In Section 2 we formulate the scattering problem in a holographic setup. We review the Polchinski-Strassler mechanism \PolchinskiUF\ for the universality of the high energy scattering at imaginary angles. When considering scattering of mesons, this picture predicts a universal correction to the Veneziano asymptotic \limitVen\ due to massive endpoints of the strings. We then use effective theory of rotating strings \HellermanKBA\ to extend the validity of this correction beyond the holographic setup. 
In Section 3 we use the worldsheet model of strings with massive ends to compute the correction to the $n$-point scattering amplitude. The dynamics of relativistic strings with massive endpoints is highly non-linear, but the leading correction to the amplitude could be computed. For the four-point amplitude with same masses at the ends, the result is \mainresult. In section 4 we explain the emergent $s \leftrightarrow u$ symmetry of the asymptotic results \limitVen\ and \mainresult. In Section 5 we analyze the result \mainresult\ using the bootstrap approach \CaronHuotICG . The new element when talking about the correction to the leading solution is the sensitivity of the asymptotic amplitude to the degeneracy of the spectrum. We further impose the emergent $s \leftrightarrow u$ symmetry and show that the massive ends correction is unique. We conclude in section 6.

\newsec{Holographic Setup}

Holography is a useful tool to study confining gauge theories, see \refs{\ErdmengerCM,\SonnenscheinPIM} for a recent review. We start this section by reviewing the pioneering work of Polchinski and Strassler \PolchinskiUF\ regarding the scattering in confining gauge theories with gravity duals. We explain how the bootstrap results of \CaronHuotICG\ follow from the holographic consideration. We then use the holographic setup to compute the first correction to the asymptotic form of the four-point amplitude at large and positive $s$ and $t$.
In this discussion, we use only the gross features of the holographic model. These features are the same in a wide range of models dual to confining gauge theories \refs{\ErdmengerCM,\SonnenscheinPIM}. Correspondingly, in the regime of interest $s,t \gg 1$ the first deviation from the Veneziano amplitude is the same in all such theories. As advertised in the introduction, this correction is a universal property of a generic theory of WIHS.

A generic holographic background dual to a confining gauge theory takes the form
\eqn\metric{
ds^2=dr^2+f(r)\,dx^2_{1,d-1} \ ,
}
where $x^\mu$ parametrizes $R^{1,d-1}$ and $r$ is a radial holographic direction. In the large $r \gg 1$ UV region the background asymptotes to AdS, $f(r)\simeq e^{2r}$. The background terminates in the IR at some finite, positive warping factor $f(r_{IR})>0$. Without loss of generality we set $r_{IR}=0$. The value of the warping factor at $r=0$ is dual to the confining scale or, equivalently, the gap in the spectrum  where the RG terminates. Hence, this is the part of the background that is relevant for us. In this region
\eqn\wrappingfactor{
f(r)\simeq1+r^2/R^2+\dots
}
where we have approximated the warping factor by its first quadratic expansion around $r=0$. For convenience, we set $f(0)=1$.

\subsec{Meson Spectrum}

To describe mesons in such a confining background, we add to it a probe flavor brane \refs{\ErdmengerCM,\SonnenscheinPIM}. 
It is a spacetime filling brane that extends from the boundary at $r = \infty$ down to some radial position $r_0 \ge 0$. Strings ending on this brane are dual to mesons in the confining gauge theory.

\ifig\confiningbackground{a) A rotating open string in the dual geometry. The vertical $r$ direction is a radial holographic direction. The open string that is dual to a meson ends on a space-filling flavor brane. When the spin of the string becomes large, it has the characteristic U-shape. The horizontal region corresponds to a string rotating in flat space, and the vertical piece of the string is dual to a constituent quark mass.  b) The picture of a meson as a rotating string with massive ends that is dual to the string depicted in a). } {\epsfxsize5.2in\epsfbox{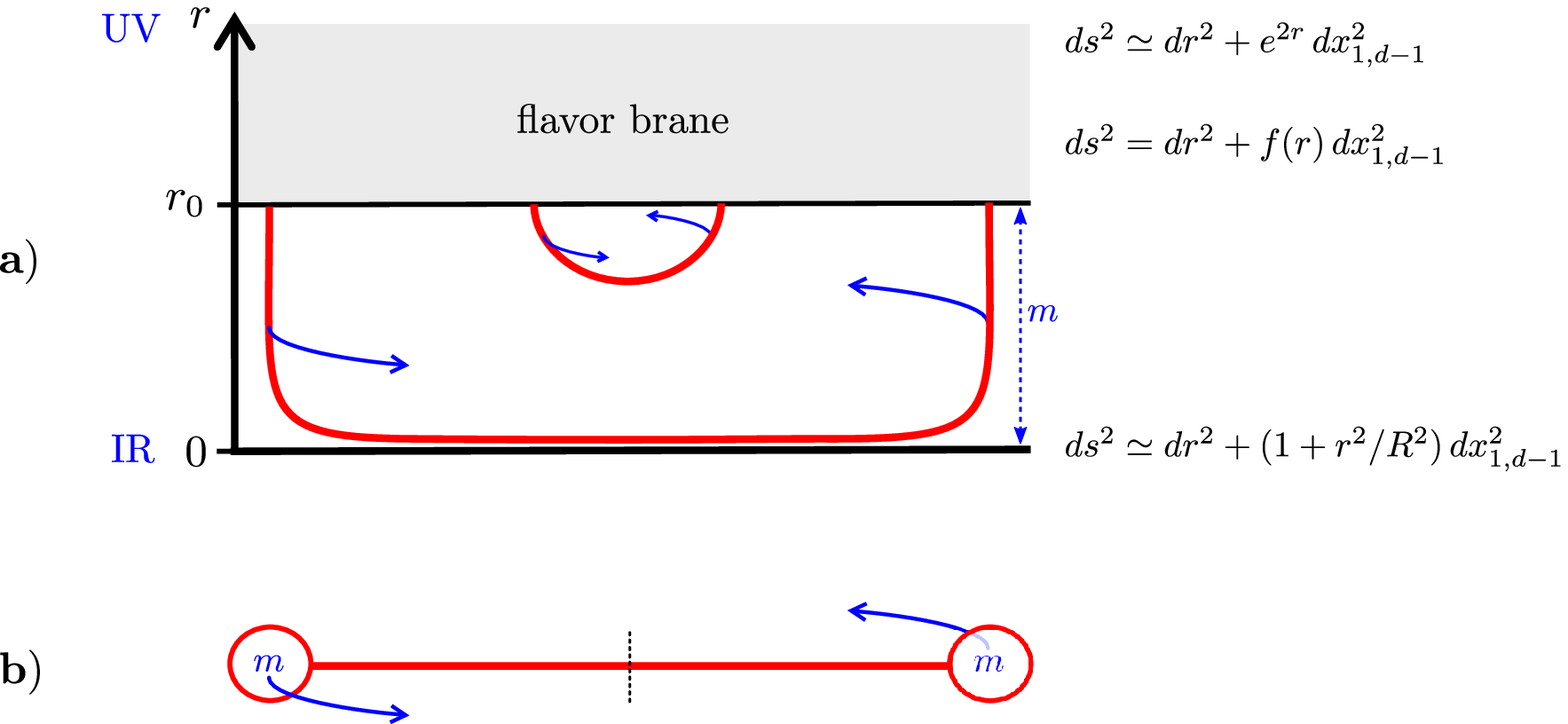}}

Mesons with low or zero spin are dual to short strings with energy of order the string scale at $r_0$, see figure \confiningbackground.a. Mesons with very large spin are dual to large rotating open strings. The main contribution to their energy is classical and can be computed by solving for the classical rotating string in the background \metric\ \KruczenskiME.

At leading order in the large spin expansion, the problem reduces to a classical rotating string in flat spacetime and the Regge trajectory is linear $J=\alpha' E^2$, where $\alpha'$ is the effective string scale in the IR. For large $r_0/l_s$, the first correction to this asymptotic trajectory is also classical and takes the form \ChodosGT
\eqn\rotating{
J=\alpha'E^2-{8\sqrt{\pi}\over3}\alpha'm^{3\over2}\sqrt{E}+{\cal O}(1)\ ,\qquad E\gg m\gg1/l_s \ ,
}
where $m=r_0/\alpha'$ .

The corresponding classical string solution has two regions, see figure \confiningbackground.a. One region is localized at $r=0$ where we have a rotating string solution in flat spacetime. In the second region, the open string extends radially between $r=0$ and the flavor brane at $r=r_0$. For large spins the string is large, and we may reliably describe it using the effective theory of long strings. In the effective theory, the second vertical piece of the string reduces to a point-like mass $m=r_0/\alpha'$ at the ends of the open string, see figure \confiningbackground.b  \KruczenskiME.

In addition to the leading order classical correction to the asymptotic trajectory \rotating, the trajectory receives a series of classical corrections as well as quantum corrections. The quantum corrections are small compared to \rotating\ when $r_0/l_s=m l_s$ is large. Hence, this is the limit we are considering here. We expect that at the next order the details of the background will become important unless some extra symmetries are present.

\subsec{Meson Scattering}

We are interested in the $2\to2$ scattering amplitude of scalar mesons in the limit $s,t\gg m^2\gg1/\alpha'$, where $\alpha'$ is the IR string scale (at $r=0$). This scattering amplitude is dual to an amplitude of open stings in the background \metric. The open strings end on the flavor brane/branes. These strings satisfy Neumann boundary condition along the $d$ flat spacetime directions. In the radial direction they satisfy Neumann boundary condition for $r>r_0$ and Dirichlet boundary condition if the string stretches from $r=r_0$ down to $r<r_0$. 

In the planar limit, the string worldsheet has a disk topology. At the boundary of the disk, we have four vertex operators that inject four asymptotic meson states with momenta $\{k_1,k_2,k_3,k_4\}$. Because these mesons have zero spin, their dual asymptotic string states are not rotating. Masses of such asymptotic states are of order the string scale. Hence, in the large $s,t$ limit that we are interested in, the four asymptotic states are approximately point-like, and their momenta are null, $k_i^2=0$. Their vertex operators take the form
 \eqn\vo{
 V_i(\sigma)=h\left(r(\sigma)\right)\times e^{ik_i\cdot x(\sigma)}}
 where $\sigma$ parametrize the boundary of the disk and $h(r)$ is some radial profile.

\subsec{Polchinski-Strassler Mechanism}

Let us first review what happens in flat space. In the Gross-Manes setup \refs{\GrossGE,\GrossKZA}, for large $|s|$ and $|t|$ the worldsheet path integral is well-approximated by a saddle point. The saddle corresponds to a large and smooth Euclidean classical string solution. There are two kinematical regimes in which the saddle differs significantly. First, there is the regime of real scattering angles. This regime was studied by Gross and Manes for open strings \GrossGE, by Gross and Mende for closed strings \GrossKZA\ and by Alday and Maldacena in a holographic setup \AldayHR. In this kinematical regime, the corresponding amplitude is exponentially small and represents a tunneling process. The other kinematical regime corresponds to purely imaginary scattering angles. In this regime, the amplitude is exponentially large.

Holographically, there is a clear difference between these two kinematical regimes. For real scattering angles, the classical solution extends towards the large $r$ UV regime of the holographic background. On the other hand, for imaginary scattering angles, the classical solution extends towards the small $r$ IR regime of the holographic background. Let us elaborate on how these very different behaviors come about:

\item{I.} When the scattering angle is real, the four external momenta are real. Without loss of generality, they can be placed in an $R^{1,2}$ subspace. On the other hand, for imaginary scattering angles, the external momenta are complex. They could also be thought of as being real in a spacetime with two times and one space, $R^{2,1}$. 
\item{II.} In either case, in the coordinates in which the external momenta are real, the classical solution is imaginary. This is because the source for these worldsheet fields is imaginary. It comes from the vertex operators and is given by $i\sum_j k_j\cdot  x(\sigma_j)$. For example, for a free string in flat spacetime we have $x^{\mu}(z,\bar z)=i\alpha'\sum_j k^{\mu}_j\log|z-\sigma_j|^2$, where $z = \sigma + i \tau$ parametrize the upper half complex plane.
\item{III.} The induced metric on the worldsheet is Euclidean. To see this it is convenient to switch to the T-dual coordinates $\partial_\alpha y^\mu=i\epsilon_{\alpha\beta}\partial_\beta x^\mu$. In these coordinates, the solution is real and ends on a null polygon made of the external null momenta \AldayHR. For example, for a free string in flat spacetime we have $y^{\mu}(z,\bar z)=\alpha'\sum_j k^{\mu}_j\log{z-\sigma_j\over \bar z-\sigma_j}$. For real scattering angles the ordering of the edges is such that every two points on two different edges of the polygon are spacelike separated. As a result, the induced metric on the worldsheet is space-space ($\partial y_\mu\bar\partial y^\mu>0$), and the Euclidean action $S_{E} >0$ is positive. Hence, the amplitude is exponentially small ($e^{-S_E}\ll1$). For imaginary scattering angles every two points on two different edges of the polygon are timelike separated and the induced metric on the worldsheet is time-time ($\partial y_\mu\bar\partial y^\mu<0$). Now, the Euclidean action is negative $S_E<0$ and the amplitude is exponentially large ($e^{-S_E}\gg1$).
\item{IV.} The action (in the conformal gauge) takes the form
\eqn\holographicac{
S_{\rm bulk}={1\over2\pi\alpha'}\int dz^2\left[\partial r\,\bar\partial r+f(r)^{-1}\partial y_\mu\,\bar\partial y^\mu\right]}
where  in the T-dual $y$-coordinates the warping factor $f(r)$ is inverted. All that will be important here is that it is a monotonically increasing function, $f'(r)>0$. The radial equation of motion that follows from \holographicac\ is
\eqn\reom{
\partial\bar\partial r=-{f'(r)\over2f(r)^2}\,\partial y_\mu\bar\partial y^\mu}
Let us consider the point in the bulk of the worldsheet where $r$ is extremal (and hence $\partial\bar\partial r=0$). The holographic coordinate $r$ changes as we go to the boundary of the disk. For real scattering angles, $\partial\bar\partial r$ is negative and hence $r$ decreases while for imaginary angles the picture is reversed.

We note that this is precisely the expected behavior that follows from unitarity \CaronHuotICG. Namely, for positive $s>0$ and $t>0$, which corresponds to the imaginary scattering angle, the amplitude is dominated by its exponentially large imaginary part. Due to unitarity, the contribution of every on-shell state is strictly positive, and there cannot be any cancellations. Hence, the amplitude is large. On the other hand, for $s>0$ and $t<0$, that correspond to real scattering angles, the cancelations are possible, and the amplitude could become small.\foot{The exponentially small result in the usual string theory requires fine-tuning and, indeed, does not hold in a generic confining gauge theory \PolchinskiTT.}

\subsec{Massive Ends Approximation} 

As a result of the discussion in the previous section, in the conformal gauge the string action reads
\eqn\action{
S={1\over2\pi\alpha'}\int dz^2\left[\d r\,\bar\d r+(1+r^2/R^2)\,\d x \cdot \bar\d x \right]+i\sum_jk_j\cdot x(\sigma_j) \ . 
}
Due to the Polchinski-Strassler mechanism, the classical string configuration for imaginary scattering angles satisfies Dirichlet boundary condition in the radial direction and Neumann in the transverse directions
\eqn\rDirichlet{
\left.\partial_\tau x^\mu\right|_{\tau=0}=i 2\pi  \alpha'   \sum_jk_j^\mu\delta(\sigma-\sigma_j)\ ,\qquad\left.r\right|_{\tau=0}=r_0  \ .
}
The equations of motion that follow from \action\ take the form
\eqn\eom{
\eqalign{R^2\,\d\bar\d r-r\,\d x_\mu\bar\d x^\mu&=0
\cr
(R^2+r^2)\d\bar\d x^\mu-r(\d r\bar\d x^\mu+\bar\d r\d x^\mu)&=i2\pi\alpha'R^2\sum_jk^\mu_j\,\delta^2(z-\sigma_j)
}}
These equations are subject to the Virasoro constraint 
\eqn\Vir{
\d r\d r+(1+r^2/R^2)\d x_\mu\d x^\mu=
\bar\d r\bar\d r+(1+r^2/R^2)\bar\d x_\mu\bar\d x^\mu=0
}

We are interested in solving these equations in the limit where $\alpha's,\alpha't\gg R/r_0\gg1$. In this limit there are three regimes, where we can solve the equations of motion:
\item{a)} Near the vertex operators insertion points $\{\sigma_j\}$. At these points the term proportional to $\d\bar\d x^\mu$ in \eom\ dominates and 
\eqn\nearvo{x^\mu(z,\bar z)\simeq i\tilde\alpha'\ k_j^\mu\log|z-\sigma_j|^2\ ,\qquad r=r_0 \  ,
}
where $\tilde\alpha'=\alpha'\left(1+r_0^2/R^2\right)$ is the string scale at $r_0$.
\item{b)} Deep inside the bulk of the string, where $r=0$ and we have the flat space solution
\eqn\flat{x^\mu(z,\bar z)=i\alpha'\sqrt{s}\left(f^\mu(z)+c.c\right) \ .}
\item{c)} In between region (a) and region (b) where $r$ varies from $r=r_0$ to $r=0$.

In the limit $s,t\gg (r_0/\alpha')^2$ we are working at one may approximate region (c) as a pointlike mass. This point particle is a domain wall between two different tensions, one at $r=0$ ($\alpha'$) and the other at $r_0$ ($\tilde\alpha'$). Its mass is given by the radial distance between the flavor brane and the IR wall
\eqn\mass{m\simeq r_0/\alpha' \ .}
After we replace region (c) by a point mass, regions (a) and (b) could be glued together. The gluing condition is equivalent to equating the difference in momentum flow between the two regions to the acceleration of the mass. This approximation may break down close to the vertex insertion. This fact, however, will not be relevant to us. The reason is that in the limit $s,t\gg m^2\gg1/\alpha'$ region (c) is pushed towards the boundary and region (a) shrinks correspondingly.

One can confirm the validity of the above approximation by working out the size and gluing of these regions in more detail. In the end, we are left with an effective description of a free string in flat spacetime with massless vertex operators insertions and a massive point particle at the boundary
\eqn\NG{
S_E^{NG}={1\over2\pi\alpha'}\int d\sigma d\tau\sqrt{\det\d_\alpha x_\mu\d_\beta x^\mu}+m\int d\sigma\sqrt{|\d_\sigma x(\sigma,0)|^2}+i\sum_jk_j^\mu x_\mu(\sigma_j) \ .
}
Here, the induced metric on the worldsheet is Euclidean, and the path integral weight is $e^{-S_E}$. Note that the point particle action is reparametrization invariant. Using this freedom, here we have already set a parametrization in which the boundary of the string is at $\tau=0$.

Above, we have written the worldsheet action back in its gauge unfixed Nambu-Goto form. The reason for that is that conformal gauge in the holographic background and the one we will use in the effective theory are not the same.\foot{This is because in the holographic background the worldsheet stress tensor also receives a contribution from the radial direction.}

\subsec{Massive Ends in The Conformal Gauge}

We now fix the conformal gauge
\eqn\virasoroflat{
(\d_\sigma x)^2-(\d_\tau x)^2=\d_\sigma x_\mu\d_\tau x^\mu=0 \ .
}
The action takes the form
\eqn\conformalgauge{
S_E^{\rm conf}={1 \over 2 \pi \alpha'} \int\! d^2 z\, \pa_z x^{\mu} \pa_{\bar z} x_{\mu}+m\int d\sigma\sqrt{|\d_\sigma x(\sigma,0)|^2}+i\sum_jk_j^\mu x_\mu(\sigma_j) \ .
}
The dynamics of a string with massive ends \conformalgauge\ is highly non-linear. Even though this model was first introduced by Chodos and Thorn in 1974 \ChodosGT\ very little regarding its properties is known, see \BarbashovNQ . One known exact solution of \conformalgauge\ which will be important to us describes the rotating string and, thus, allows one to compute the correction to the Regge trajectory. In analyzing this model we find it convenient to rewrite the worldline action using the boundary metric\foot{Some signs here may not look standard. This is because here we are considering a Euclidean solution.}
\eqn\action{
S = {1 \over 2 \pi \alpha'} \int\! d^2 z\, \pa_z x^{\mu} \pa_{\bar z} x_{\mu} + {1 \over 2} \int d \sigma \left( e\,\pa_\sigma x^{\mu} \pa_{\sigma} x_{\mu} + {m^2 \over e}\right) + i \sum_j k_j^{\mu} x_{\mu} (\sigma_j) \ ,
}
where
\eqn\boundarycond{
e(\sigma)^2={m^2\over \d_\sigma x^\mu\d_\sigma x_\mu}\ .
}
As we explain below introducing $e(\sigma)$ makes perturbative expansion in $m$ very natural. Evaluation of the amplitude in terms of $e(\sigma)$ will turn out very natural as well. 

To summarize, in the limit $s,t\gg m^2\gg1/\alpha'$ the amplitude is given by 
\eqn\summary{\log A(s,t)\simeq -S_{\rm classical}}
where $S_{\rm classical}$ is the extremum of the action \action. The first correction to this saddle point due to the massive endpoints is universal because we have only used the gross features of the holographic model which are shared by all models dual to a confining gauge theory.  

It turns out that this effective description is also valid when the quantum corrections are not small compared to the (bare) boundary mass.  Namely, \summary\ also holds when $m^2\alpha'$ is not large, but the numerical value of $m^2\alpha'$ may change. Even more than that, we expect \summary\ to hold for closed strings too. To understand why it is so we will now discuss this correction using the effective theory of long strings.

\subsec{Effective Theory of Long Strings}

One can systematically study all possible corrections to the dynamics of very long rotating strings. Understanding these corrections is the subject of the effective theory of long strings \PolchinskiAX.
The leading order result in this expansion is fixed by symmetry and is given by the Nambu-Goto action. The spectrum of long rotating open strings was studied using the effective theory in \HellermanKBA.\foot{For the long static string, the set of universal corrections was classified in \AharonyIPA.} In this study, one starts from the free long rotating string and considers all possible higher derivative operators that are consistent with the symmetries of the problem. There are two types of corrections -- due to the boundary and the bulk operators. 

The first operator one can add on the boundary has the effect of adding mass at the endpoint of the string. As discussed above, it leads to a $\sqrt E$ correction to the asymptotic trajectory \rotating. There are no other boundary corrections to the trajectory that grow with energy. 

The bulk operator that gives the leading correction to the asymptotic trajectory is the so-called Polchinski-Strominger term \PolchinskiAX
\eqn\PS{
{\cal L}_{\rm PS}=\beta{(\partial^2x^\mu\bar\partial x_\mu)(\partial x^\nu\bar\partial^2 x_\nu)\over(\partial x^\mu\bar\partial x_\mu)^2} \ .
}
In the holographic scenarios, this operator is generated by integrating out quantum fluctuations to the quadratic order. It turns out that the Polchinski-Strominger term \PS\ is singular at the boundary of the rotating string \HellermanKBA . As a result, it induces a boundary operator that has the effect of introducing a boundary mass.\foot{In \HellermanKBA\ this mass was fine-tuned to zero.} Hence, we are back in the massive endpoints case \rotating. After canceling the boundary divergence, the remaining bulk contribution is regular and only lead to power suppressed corrections to the asymptotic trajectory. 
\ifig\general{a) We imagine quantum corrections on the worldsheet. For the rotating string, they experience the centrifugal potential and are pushed to the boundary of the string. The leading effect is making the endpoints of the string massive. b) Imagine a rotating closed string. It has folds, points where the extrinsic curvature diverges. Regularizing it by adding higher derivative terms in the worldsheet action makes the endpoints move slower than the speed of light. The result is again the string with the massive endpoints.} {\epsfxsize5.2in\epsfbox{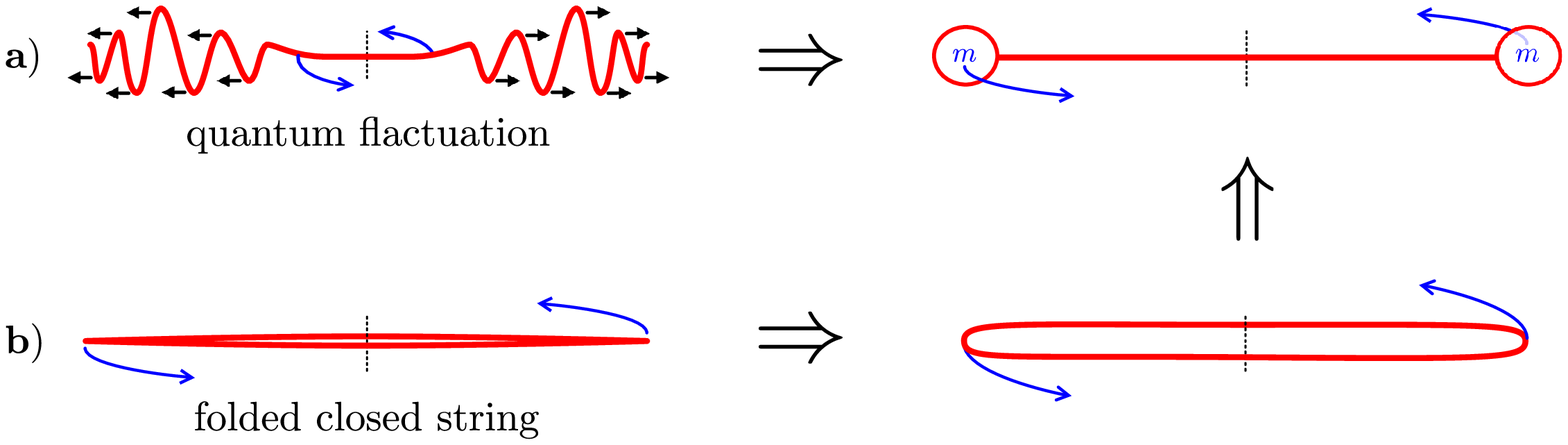}}

There is an intuitive way to understand this effect. Suppose we start with no bare boundary mass ($r_0=0$) and consider a quantum fluctuation of the string in directions transverse to the rotating plane. This fluctuation is then pushed by the centrifugal force towards the boundary, see figure \general.a. Since it carries energy, it has the same leading order effect as adding a boundary mass.\foot{We thank Sergei Dubovsky for explaining this point  to us.} 

To summarize -- the leading correction to the asymptotic linear trajectory comes from slowing down the boundary endpoints of the rotating string so that they are no longer moving at the speed of light. As a result, the asymptotic trajectory receives a $E^{1/2}$ correction. The sign of this correction is negative because the end of the string slows down so that it carries less angular momentum.

Let us briefly discuss the case of closed strings. The free closed rotating string has two tips where the extrinsic curvature diverges. Similar to the open string case for the rotation in one plane the Polchinski-Strominger term diverges. As a result, the effective theory breaks down, and the tools used in \HellermanKBA\ become inapplicable. That is why we leave the consideration of closed string to the future and only present here our intuitive expectation. Imagine we regularize the cusps divergences in some generic way. After regularization, (and without fine-tuning), the curvature will be everywhere finite. As a result, the tips of the rotating string will be smoothed out, see figure \general.b. Such tips carry some energy and can no longer move at the speed of light. So again, the leading effect at large spin is to slow down the tips, effectively adding there a mass. That is why we expect the correction to be much more general than what we have shown here. One hope is to be able to address this issue using the bootstrap. We do that in the last section of the paper.

So far we have summarized the results of the effective theory for the spectrum of large rotating strings. In this paper, however, we are interested in the scattering problem. When studying a scattering process, one cannot assume a priori that we have a long string plus small corrections. Indeed, as follows from the Polchinski-Strassler consideration, the long string description is only valid for large and positive $s,t$. The effective theory does not have any way of knowing that and this is partly the reason we did not start with this point of view. On the other hand, the effective theory allows us to extend the validity of the massive endpoints model beyond what one may deduct from the holographic setup.

\newsec{Classical Scattering of Strings With Massive Ends}

We shall now use the effective massive endpoints model \action\ to compute the first non-trivial correction to the asymptotic Veneziano amplitude. This computation amounts to extremizing the action \action\ with respect to $x^\mu$, $e$, $\{\sigma_j\}$ and then evaluating the action on-shell, \summary. Solving the corresponding non-linear equations of motion is hard. However, we will only need to solve them to the first non-trivial order in the mass. As discussed above, higher orders are non-universal and will not be of interest to us.

Let us start by listing the equations of motion we would like to solve. First, we have the boundary conditions for $x$. Due to the mass at the boundary of the string, the usual Neumann boundary condition is modified to
\eqn\bdry{
{1 \over 2 \pi \alpha'} \pa_\tau x^{\mu} +  m\,\pa_{\sigma}{\pa_{\sigma} x^{\mu}\over\sqrt{\d_\sigma x^\nu\,\d_\sigma x_\nu}}
=i \sum_{j} k_j^{\mu}\, \delta(\sigma - \sigma_j) \ .
}
This equation has a simple interpretation. The momentum flow from the bulk of the string is equal to the acceleration of the mass at the boundary.  In addition, the vertex operators inject momenta $k_j$ at $\sigma=\sigma_j$. 

In the bulk of the string, $x^\mu$ is a free field satisfying 
\eqn\bulkeom{
\d\bar\d\, x^\mu(z,\bar z)=0\qquad{\rm for}\qquad \tau=(z+\bar z)/2>0 \ .
}

In addition to the worldsheet field $x^\mu$, we have to minimize the action with respect to the vertex operators insertion points $\{\sigma_j\}$. Doing so is equivalent to imposing the Virasoro constraints
\eqn\Virasoro{
\pa x^{\mu}\,\pa x_{\mu}=\bar \pa x^{\mu}\,\bar\pa x_{\mu} = 0 \ .
}

Naively, we have a standard perturbative expansion in integer powers of the mass. However, the main difficulty in the calculation comes from the fact that the naive small $m$ expansion is singular. In the absence of the mass, the ends of the string move at the speed of light. Hence, even though the second term in the LHS of \bdry\ is proportional to the mass, it is inversely proportional to the boundary velocity of the string ${1 \over \sqrt{1 - v^2}}$ which is infinite when $v=1$.  As a result, the expansion re-organizes itself in powers of $\sqrt{m}$. 

To see how this fractional scaling emerges, we first split $x^\mu$ as
\eqn\exact{
x^{\mu} = x_{0}^{\mu} + y^{\mu}.
}
where $x_0^\mu$ is the solution with no boundary mass. It satisfies Neumann boundary conditions, (away from the insertion points), and is given by\foot{Recall that $\pa_{z} \pa_{\bar z} \log z \bar z = 2 \pi \delta^{(2)}(z,\bar z)$.} 
\eqn\solutionnew{
x_{0}^{\mu}(z,\bar z)  = i \alpha'  \sum_i k_i^{\mu} \log |z-\sigma_i|^2
}
where at this point, the $\sigma_j$'s are undetermined. Now, the boundary tangent is given by
\eqn\bvelocity{
\d_\sigma x \cdot \d_\sigma x=\d_\tau x \cdot \d_\tau x=\d_\tau  y \cdot \d_\tau y  \qquad{\rm for}\qquad\tau=0\ ,\quad\sigma\ne\sigma_j
}
where in the first step we have used the Virasoro constraint \Virasoro. By plugging the boundary tangent \bvelocity\ into the the boundary conditions for $x^\mu$, \bdry, we learn that
\eqn\yscaling{
y(z,\bar z)\propto\sqrt{m}
}
plus higher order corrections. 

In what follows, we will first derive a general expression for the correction to the amplitude in terms of the leading order solution. In particular, our result holds for an arbitrary number of particles. We will then focus on the case of the four-point amplitude. It would be convenient to use the worldline metric \boundarycond\ that also scales as $e(\sigma)\propto\sqrt{m}$.

\subsec{Solving for the Boundary Metric}

We first note that the perturbation $y^\mu$ is also a free field in the bulk of the string. Namely, $\d\bar\d\,y^\mu(z,\bar z)=0$. At the boundary, it satisfies the boundary condition
\eqn\bdry{
\pa_\tau y^{\mu} =-2 \pi \alpha'\,\pa_{\sigma} (e\, \pa_{\sigma} [x_0^{\mu} + y^{\mu}]).
}
We may think about the right hand side as a continuous momentum source for $y^{\mu}$. Hence, similarly to \solutionnew, we can express the correction $y^{\mu}$ as
\eqn\exact{\eqalign{
y^{\mu}(z, \bar z) &=-\alpha'\int\! d\sigma\,\pa_{\sigma} (e\, \pa_{\sigma} [x_0^{\mu} + y^{\mu}])\log|z-\sigma|^2\cr
&= \alpha'  \int\! d\sigma\, e\,\pa_\sigma[x_0^{\mu} + y^{\mu}]  \left( {1 \over \sigma - z} + {1 \over \sigma - \bar z} \right) \ ,
}}
where in the second step we have preformed an integration by parts. Here, $z$ ($\bar z$) takes values in the upper (lower) half-plane.

Next we can use \exact\ together with the expression for the boundary velocity \bvelocity, to get the following equation for $e(\sigma)$
\eqn\equationfinitecoubling{\eqalign{
{1 \over (2 \pi \alpha')^2}{m^2 \over e^2}& ={1 \over (2 \pi \alpha')^2}\d_\sigma x^\mu\d_\sigma x_\mu \cr
&=e^2  \pa_{\sigma}^2 x_0^{\mu} \pa_{\sigma}^2 x_0^{\mu}  - m^2 \left[\pa_\sigma \log e\right]^2 + e^2 \left[ 2 \pa_{\sigma}^2 x_0^{\mu} \pa_{\sigma}^2 y^{\mu} +  \pa_{\sigma}^2 y^{\mu} \pa_{\sigma}^2 y^{\mu} \right] .
}}

So far, we have not done any small mass approximation. That is, \exact\ and \equationfinitecoubling\ are exact equations, valid for any value of $m$. Next, we note that the leading order correction corresponds to 
\eqn\escaling{
e(\sigma) \propto\sqrt{m}.
}
In this way, the first term in the RHS of \equationfinitecoubling\ is of order $m$, while other two terms are of order $m^{3/2}$ and $m^2$ correspondingly. Therefore, we get that the first correction to the boundary metric is
\eqn\LOsolution{
e_*(\sigma)^2={m\over2 \pi \alpha'\sqrt{\pa_{\sigma}^2x_0\cdot\pa_{\sigma}^2x_0}} ,
} 
where $x_0^{\mu}$ here is simply the usual Gross-Mende solution \solutionnew\ and the $\sigma_j$'s being the solution of scattering equations
\eqn\scatteringequations{
\sum_j {k_i\cdot k_j \over \sigma_i - \sigma_j } = 0
} 
for every $i= 1, ... , n$  \GrossKZA.

The expression for the boundary metric \scatteringequations\ is general. In particular, it holds for an arbitrary number of particles. Once we restrict to the case of the four-particle amplitude, we find that
\eqn\efourpoint{
e_{n=4}(\sigma)=c\,\sqrt{m}\prod_{j=1}^4\sqrt{|\sigma-\sigma_j|}\qquad{\rm where}\qquad c^{-4}=(4\pi)^2\alpha'^4(s+t)\prod_{j=1}^4|\sigma_{j+1}-\sigma_j|
}
Importantly, we see that the induced metric vanishes at the vertex operators insertion points. This gauge independent property comes hand in hand with the fact that at these points we inject null momenta into the string.

Next, we turn to the evaluation of the on-shell action.

\subsec{Correction to the Amplitude}

To compute the correction to the amplitude we need to evaluate the on-shell action \action\ on the classical solution for $x^{\mu}$, $e$ and the insertion points $\{\sigma_j\}$. At the leading order, $m=0$, we have the Gross-Mende (GM) solution \solutionnew, (or more precisely, its analog for imaginary scattering angles). The on-shell action is given by one half times the value of the source, evaluated on-shell
\eqn\usualGM{
S_{GM}=  {i \over 2}\sum_jk_j\cdot x_0(\sigma_j) \ 
}
where the $\sigma_j$'s solve the scattering equations \scatteringequations. For the four-point amplitude, \usualGM\ evaluates to \limitVen.

For the correction, the situation is similar. We first subtract from \action\ the leading GM order \usualGM. We remain with an action for the correction. It is a functional of $y^{\mu}$, $e$ and the corresponding shifts in the vertex operators insertion points $\{\sigma_j\}\to\{\sigma_j+\delta_j\}$. That is,
\eqn\correctionaction{
\delta S[m;e,y^{\mu} ,\{\delta_j\}]=S-S_{GM}
}
Now the source in $\delta S$ is the mass term in \action, $\int d\sigma\, {m^2\over e}$. The rest of the terms in the action are functionals of $y^{\mu}$, $e$ and the $\delta_j$'s, but not of $m$. Therefore, $y^{\mu}$ and the $\delta_j$'s only depend on $m$ through their interaction with $e$. After solving for them in terms of $e$, we remain with a functional of $e$ only. The details of this are not important, all we need is that $y^{\mu}$ and the $\delta_j$'s are determined by $e$ (and of course, the zeroth order solution). We have
\eqn\deltaS{
\delta S[m;e]=\delta S\left[m;e, y^{\mu}[e],\{\delta_j[e]\}\right]
}

In appendix A we have calculated $\delta S[m;e]$ explicitly for the case of four external particles. We know that the correction scales as $\delta S[m;e]\propto m^{3/2}$ and $e\propto\sqrt m$. Therefore,
\eqn\cubic{
\delta S[m;e]={1\over2}\int d\sigma\, {m^2\over e(\sigma)}+\int\! d\sigma_1\,d\sigma_2\,d\sigma_3\,e(\sigma_1)\,e(\sigma_2)\,e(\sigma_3)\, \Lambda(\sigma_1,\sigma_2,\sigma_3)
}
for some function $K$. We can now apply an integrated version of the virial theorem. Namely, we have
\eqn\virialint{\eqalign{
0&=\left.\int d\sigma\,e(\sigma){\delta\over\delta e(\sigma)}\delta S[m;e]\right|_{e=e_*}\cr
&=-{1\over2}\int d\sigma\, {m^2\over e_*(\sigma)}+3\int\! d\sigma_1\,d\sigma_2\,d\sigma_3\,e_*(\sigma_1)\,e_*(\sigma_2)\,e_*(\sigma_3)\,\Lambda (\sigma_1,\sigma_2,\sigma_3)
}}
We conclude that (see appendix A for a detailed derivation of this result)
\eqn\onshellaction{
S_{on-shell} = S_{GM} + {2 \over 3} \int d \sigma {m^2 \over e_*(\sigma)}= S_{GM} + {2 \over 3}\sqrt{2\pi\alpha'}\,m^{3/2}\int d \sigma\left(\d_\sigma^2x_0\cdot\pa_{\sigma}^2x_0\right)^{1/4}\ .
}

Note that even though this correction depends on the second derivative of $x_0$, it is reparametrization invariant. The reason for this is that before we turn on the mass, the endpoints of the open string moves at the speed of light and hence $\d_\sigma x_0\cdot \d_\sigma^2 x_0=0$.

Equation \onshellaction\ is the main result of this section. It is the first correction to any classical string configuration due to a small mass at the endpoints of the string. Once we plug in  \onshellaction\ the GM solution for four external particles we get
\eqn\mainagain{\eqalign{
\log A(s,t)=-S_{on-shell}&= \alpha' \left[ (s+t) \log (s+t) - s \log s - t \log t \right] \cr
&- {16 \sqrt{\pi} \alpha' \over 3}m^{3/2} \left( {s t \over s + t} \right)^{1/4} \left[K\left({s \over s + t} \right) + K\left({t \over s + t} \right) \right] + ... \ .
}}

Let us take the Regge limit of this result $s \gg t$. The amplitude takes the standard form $\log A(s,t) = j(t) \log s$ with the Regge trajectory being
\eqn\Reggecorr{
j(t) = \alpha'  \left(t -  {8 \sqrt{\pi} \over 3} m^{3/2} t^{1/4} + ... \right) .
}
This correction to the Regge trajectory matches precisely with the one obtained before for the explicit rotating string solution, see for example formula (87) in the review \SonnenscheinPIM. 

For us, the correction in \Reggecorr\ comes from the first elliptic function in \mainagain. Namely, the first elliptic function captures the correction to the spectrum in the $t$-channel and the second elliptic function captures the correction to the spectrum in the $s$-channel. It turns out that this have a good physical reason. Going back to the holographic setup, we may consider several flavor branes instead of one. Each flavor brane can end at its independent radial position, $r_{0,j}$. We can now scatter the open strings that stretch between different types of flavor branes. Correspondingly, we will end in the effective theory with different mass between different vertex operators insertions. We label the mass between vertices with momentum $k_j$, $k_{j+1}$ by $m_{j,j+1}$. The masses $m_{2,3}$ and $m_{4,1}$ correct the spectrum in the $s$-channel and the masses $m_{1,2}$ and $m_{3,4}$ correct the spectrum in the $t$-channel. The correction to the amplitude now reads
\eqn\differentmasses{
\delta S[m;e_*]={8 \sqrt{\pi} \alpha' \over 3} \left( {s t \over s + t} \right)^{1\over4} \left[\left(m^{3/2}_{1,2}+m^{3/2}_{3,4}\right) K\left({s \over s + t} \right) + \left(m^{3/2}_{4,1}+m^{3/2}_{2,3}\right) K\left({t \over s + t} \right) \right].
}

 Similarly, for the $n$-point amplitude we have
\eqn\differentmassesn{
\delta S[m;e_*]={2 \over 3}\sqrt{2\pi\alpha'}\ \sum_{i} \ m_{i,i+1}^{3/2}\int\limits_{\sigma_i}^{\sigma_{i+1}} d \sigma\left(\d_\sigma^2x_0\cdot\pa_{\sigma}^2x_0\right)^{1/4} \ .
}

\newsec{Discontinuity, Lorentzian Segments and Emergence of The $s-u$ Crossing}

Consider an amplitude where external particles that transform in the bi-fundamental representation of a flavor symmetry, so that their corresponding partial amplitude takes the form ${\rm Tr}(T_1T_2T_3T_4)A(1, 2 ,3 , 4)$. The three Mandelstam invariants are
\eqn\Mandelstaminv{
s=-(k_1+k_2)^2\ ,\qquad t=-(k_1+k_4)^2\qquad{\rm and}\qquad u=-(k_1+k_3)^2 \ .
}

Such an amplitude is crossing symmetric with respect to the $s$- and $t$-channels ($A(s,t) = A(t,s)$), but not the $s$- and $u$- channels ($A(s,t) \neq A(u,t)$).\foot{To make it concrete the reader can think of the Veneziano amplitude $A(s,t) = {\Gamma(-s)\Gamma(-t) \over \Gamma(-s-t)}$.} 

Still, the asymptotic of the amplitude, both \limitVen\ and \differentmasses, is symmetric under the exchange of $s$ and $s \to u=-s-t$.\foot{Since we are in the regime $s,t\gg 1$ we can neglect the external masses.} More precisely, if we analytically continue $s\to u$, the real part of the amplitude \limitVen\ stays intact. The same is true about the correction \differentmasses . We can formulate this property as a vanishing of the double discontinuity
\eqn\ddis{
{\rm dDisc}_s\log A(s,t)\equiv \log A(-s-t+ i \eps, t) + \log A(-s-t - i \eps, t) - 2 \log A(s,t) =0 \ .
}

\ifig\cuttingandgluing{Analytic continuation that connects the $s$- and $u$-channels. a) The original worldsheet with ordered external states $\left<\Tr\, V_1(k_1)V_2(k_2)V_3(k_3)V_4(k_4)\right>$. b) The worldsheet after the analytic continuation $s\to u$. This new classical solution can be obtained from the amplitude with ordering of points $2$ and $3$ is interchanged $\left<\Tr\,V_1(k_1)V^\intercal_3(k_2)V^\intercal_2(k_3)V_4(k_4)\right>$ in two steps. First, we cut it in the middle of the $t$-channel and then insert a piece of the Lorentzian evolution. This Lorentzian piece contributes an imaginary piece to the Euclidean action. Its sign depends on the way we analytically continue $s\to u$.  } {\epsfxsize6in\epsfbox{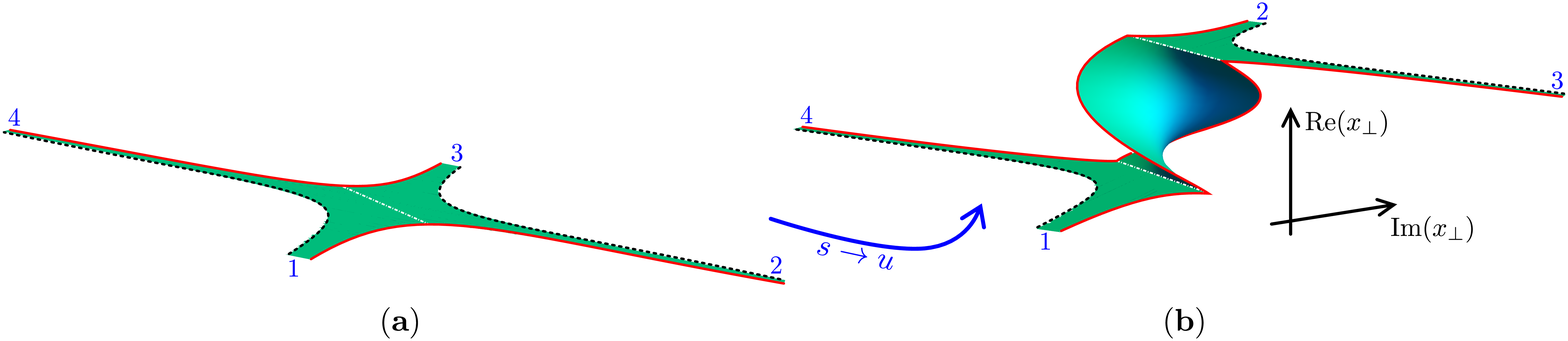}}

This is an emergent symmetry and it calls for an explanation. The purpose of this section is to provide such an explanation. As we will see it involves several ingredients:

\item{I.} The analytic continuation of the saddle point could be interpreted as an insertion of the piece of the string that evolves in the imaginary worldsheet Euclidean time, see vertical segment in \cuttingandgluing b.  The on-shell action on this Lorentzian solution is equal to the imaginary part of the analytically continued asymptotic amplitude ${\rm Im}[ \log A(-s-t,t)] = i S_{Lorentzian}$.\foot{We thank Shota Kumatso for useful discussions on related topics.}

\item{II.} The effect of this Lorentzian evolution is two-fold, see \cuttingandgluing b . First, the momenta $k_2$ and $k_3$ are mapped one into each other. Second, if we project to the Euclidean pieces of the solution, it looks as their correspondent vertex operators are transposed and their order is interchanged. Due to these effects, the Euclidean saddle maps into itself, $\left<\Tr\, V_1(k_1),V_2(k_2),V_3(k_3),V_4(k_4)\right>=\left<\Tr\,V_1(k_1)V^\intercal_3(k_2)V^\intercal_2(k_3)V_4(k_4)\right>$ and the ${\rm Re}[ \log A(-s-t,t)]$ which is the on-shell action on the analytically continued Euclidean solution is the same as the original one.

\item{III.} The fact that as we analytically continue the vertex operators get interchanged and transposed is a kind of Stokes phenomenon. Indeed, if we are to do the analytic continuation of the full amplitude $A(s,t)$ before taking the limit, we would get $\left<\Tr\, V_1(k_1)V_2(k_3)V_3(k_2)V_4(k_4)\right>\ne\left<\Tr\, V_1(k_1)V_2(k_2)V_3(k_3)V_4(k_4)\right>$. 

The punchline is that the emergent symmetry \ddis\ is a manifestation of some kind of stringy Stokes phenomenon which turns an analytic continuation to an a priori different physical regime into a symmetry transformation. 

The fact that after the continuation to the Lorentzian time we end up with a periodic solution is non-trivial. In the full microscopic theory, we have a combination of all states propagating in the $t$-channel and there is no reason for this superposition of eigenstates to have a periodicity in the Lorentzian ``time." This periodicity emerges when we look at the logarithm of the amplitude in the classical limit $s,t \gg 1$. 
Similarly, the fact that the real part of the result is $s\leftrightarrow u$ invariant does not hold at the full microscopic level and is an emergent symmetry of the classical limit $s,t \gg 1$.

\subsec{Derivation}

Let us now see how this mechanism comes about in detail. We start with the explicit free string classical solution \solutionnew\ and follow the analytic continuation. 
We then argue that any boundary correction cannot affect the result. Finally, we verify that the Lorentzian piece of the solution indeed matches with the imaginary part of the analytically continued corrected amplitude \mainagain.

For convenience, we work in the conformal gauge, with $z=e^{\tau+i\sigma}$ parameterizing the upper half-plane. We fix the frame both on the worldsheet and in spacetime. In spacetime we go to the center of mass frame, where $\vec k_1+\vec k_2=0$. On the boundary of the worldsheet, we have four vertex operators located at $\{z_1,z_2,z_3,z_4\}$. We do an $SL(2)$ transformation to place these vertex operators at
\eqn\worldsheetframe{
\{z_1,z_2,z_3,z_4\}=\{-e^a,-e^{-a},e^{-a},e^a\}\qquad{\rm where}\qquad {z_{1,2}z_{3,4}\over z_{1,4}z_{2,3}}=\sinh^2(a)=s/t \ .
}
With this choice, the energy flow across any point on the slice at $|z|=1$ is zero. 
We choose to analytically continue $s\to s^\circlearrowleft= -s-t$ in the upper complex $s$ plane.\foot{If we instead analytically continue $s\to s^\circlearrowright= -s-t$ in the lower complex $s$ plane, then we get that $a^\circlearrowright=a-i{\pi\over2}$. 
} After this analytic continuation, we arrive at
\eqn\ccrafterac{
\sinh^2(a^\circlearrowleft)=u/t=-1-\sinh^2(a)\ ,\qquad a^\circlearrowleft=a+i{\pi\over2}
}
Correspondingly, we have
\eqn\worldsheetframeafter{
\{z_1^\circlearrowleft,z_2^\circlearrowleft,z_3^\circlearrowleft,z_4^\circlearrowleft\}=\{-e^{a+i{\pi\over2}},-e^{-a-i{\pi\over2}},e^{-a-i{\pi\over2}},e^{a+i{\pi\over2}}\}\propto \{-e^a,-e^{-a-i\pi},e^{-a-i\pi},e^a\}
}
where in the last step we did an overall resealing all the insertion points. After the analytic continuation we have that $z_1^\circlearrowleft<z_3^\circlearrowleft<z_2^\circlearrowleft<z_4^\circlearrowleft$, so it seems as if the vertex operators at $z_2$ and $z_3$ have interchanged their order. However, as we explain next, the contour of integration in the complex $\tau$-plane respects the original ordering. 

At this point we only have specified the location of four vertex operators in the $\tau$-plane. We should now specify the rest of the string boundary. To fix it we impose that the string satisfies Neumann boundary conditions. These are satisfied at $\sigma=0$ and $\sigma=\pi$, with arbitrary $\tau$, (recall that $z=e^{\tau+i\sigma}$ and $\bar z=e^{\tau-i\sigma}$ weather $\tau$ is real or not). To have a clear separation into the Euclidean and Lorentzian regions we choose the boundary as follows
\eqn\newboundary{
\sigma\in\{0,\pi\}\ ,\qquad \tau\in\{[-\infty-i\pi,\tau_*-i\pi],[\tau_*-i\pi,\tau_*],[\tau_*,\infty]\}
}
where $-a<\tau_*<a$. 
In addition, we choose $\tau_*=0$. (As discussed above, at any point along this slice the energy flow between the two sides of the string vanishes.)

By letting $\sigma$ to vary continuously between $\sigma=0$ and $\sigma=\pi$, this new boundary curve continues nicely into the bulk of the string. Correspondingly, depending on whether we analytically continue $s\to u$ through the upper or lower complex plane, we end with the new classical string solutions
\eqn\newxzero{\eqalign{
x_0^\circlearrowleft&=x_0(e^{\tau+i\sigma},e^{\tau-i\sigma})\qquad{\rm with}\qquad \sigma\in[0,\pi]\ ,\ \  \tau\in\{[-\infty-i\pi,-i\pi],[-i\pi,0],[0,\infty]\}\cr
x_0^\circlearrowright&=x_0(e^{\tau+i\sigma},e^{\tau-i\sigma})\qquad{\rm with}\qquad \sigma\in[0,\pi]\ ,\ \  \tau\in\{[-\infty+i\pi,+i\pi],[+i\pi,0],[0,\infty]\}}
}

The string solution $x_0^\circlearrowleft$ has three regions discussed above and is plotted in figure \cuttingandgluing.b. The Lorentzian region corresponds to the piece where $\tau\in[0,i\pi]$. The two Euclidean regions can be glued into the original solution, (where $\{V_2(k_2),V_3(k_3)\}$ is interchanged with $\{V_3^\intercal(k_2),V_2^\intercal(k_3)\}$). This is evident once we plug \worldsheetframeafter\ into \solutionnew\ and noticing that
\eqn\reflection{
x_0^{\mu}(e^{\tau+i\sigma},e^{\tau-i\sigma})=x_0^{\mu}(e^{(\tau+i\pi)+i(\pi-\sigma)},e^{(\tau+i\pi)-i(\pi-\sigma)}) + {\rm const}\ .
}
Hence, the real part of the logarithm of the amplitude is equal to the amplitude with $s$ and $u$ interchanged. It remains to evaluate the action on the Lorentzian region. We find that
\eqn\impiece{
{1 \over 2 \pi \alpha'} \int\limits_0^{i\pi}d\tau\int\limits_0^\pi d\sigma\left[\left(\pa_\tau x_0(e^{\tau+i\sigma},e^{\tau-i\sigma})\right)^2+\left(\pa_\sigma x_0(e^{\tau+i\sigma},e^{\tau-i\sigma})\right)^2\right]=i\pi\alpha'\, t
}
in agreement with the discontinuity of \limitVen.

Next, we turn on the massive endpoints correction. Our result for the correction to the action \onshellaction\ was general, for any classical solution and any parametrization of the boundary. Hence, we just need to evaluate it on the two new boundary segments at $\sigma=0$ and at $\sigma=\pi$ with $\tau\in[0,i\pi]$. We find that
\eqn\masspiece{
\int\limits_0^{i\pi} d \tau\left[\d_\tau^2x_0(e^{\tau},e^{\tau})\cdot\pa_{\tau}^2x_0(e^{\tau},e^{\tau})\right]^{1/4}
=2 \sqrt{2}[s\,t(s+t)]^{1\over4} {K({t\over s+t })\over\sqrt{s+t}} \,
}
This is precisely the imaginary part of the analytic continuation of the corresponding term $ \left( {s\, t \over s + t} \right)^{1\over4} K\left({s \over s + t} \right)$ in \differentmasses. Note that only the $m_{1,2}^2$ and $m_{3,4}^2$ boundaries contribute to the imaginary part.\foot{In the discussion above we assumed that $m_{1,2}^2=m_{3,4}^2$ and $m_{2,3}^2=m_{4,1}^2$. This restriction was used to simplify the discussion but is not necessary.} Indeed, the other piece $ \left( {s \, t \over s + t} \right)^{1\over4} K\left({t \over s + t} \right)$ maps to itself under $s \to -s-t$. This explains why the double discontinuity of the mass correction had to vanish.

The property of the classical solution that its double discontinuity vanishes is not bounded to the first correction. It is a general property valid for any boundary interaction. The reason is that no matter what the boundary interaction is, the bulk of the string is free. In the frame where the four vertex operators are at
\eqn\fourveteces{
\{z_1,z_2,z_3,z_4\}=\{-e^a,-e^{-a},e^{-a},e^a\}
}
the solution is a linear combinations of terms of the form $\log|z-z_*|^2$, sourced at the boundary points $z_*$ and $-z_*$. As before, under the analytic continuation  $\tau\to\tau\pm i\pi$, this function goes back to itself plus a constant shift. Moreover, after the analytic continuation \worldsheetframeafter\ the solution goes to the one where vertices $V_2$ and $V_3$ are interchanged and the details are exactly the same as above. To find the analytic continuation of the vertex insertion points, parametrized by $a$, we have used their relation to the Mandelstam invariants \worldsheetframe. Even though this relation is now modified, the analytic continuation of $a$ stays the same. To see that we note that after the continuation $a^\circlearrowleft=a+i{\pi\over2}$, the solution still extremize the real part of the action and looks as the original saddle point problem with $k_2$ and $k_3$.

This picture breaks down when we start including interactions in the bulk of the string. However, since the classical string is of the size $\sqrt s$, in the large $s,t$ limit all such bulk corrections are power suppressed.

To summarize the vanishing of the double discontinuity \ddis\ is a stringy property that emerges in the classical large $s,t$ limit. We will use it in our discussion of the bootstrap below.

\newsec{Bootstrap}

In the previous sections, we started from a theory of strings and derived the first correction to the asymptotic form of the amplitude. Hence, both the leading contribution and the correction are fingerprints of the string. In this section, we take an opposite point of view and do not assume a theory of strings. Instead, we analyze the computation of previous sections using the asymptotic bootstrap initiated in \CaronHuotICG. One of the goals of this bootstrap program is to classify all possible theories of weakly interacting higher spin particles (WIHS) using unitarity and causality. Comparing the asymptotic form of the amplitude in a general theory of WIHS to the one obtained above is an indication to what extent all such theories are theories of strings.

We start by reviewing our definition of theories of WIHS. We summarize the main ideas of the bootstrap program of \CaronHuotICG\ and the argument that fixes the leading asymptotic of the amplitude to be the one of Veneziano. Next, we turn to the discussion of corrections. We check that the correction we found using the string with massive ends model fits perfectly into bootstrap expectations. The main novelty in the discussion of corrections is sensitivity to the microscopic degeneracy of the spectrum. In particular, the massive ends correction corresponds to a non-degenerate microscopic spectrum of particles (this should be juxtaposed with the Veneziano amplitude whose spectrum is maximally degenerate). 

As explained in the previous section, both the leading solution and the massive ends correction have an extra property of the $s-u$ crossing or zero double discontinuity \ddis. For scattering of non-identical particles, this property is the manifestation of a peculiar Stokes phenomenon. For scattering of identical particles, this property means that there is no Stokes phenomenon as we analytically continue $s\to u$ for imaginary scattering angles. We do not have an independent bootstrap argument that it has to be the case. Still, at the end of this section, we utilize the condition that the double discontinuity vanishes \ddis\ in the bootstrap approach. There, it translates into an integral equation on the density of the excess zeros. We show analytically that this integral equation does not have a solution unless the correction to the asymptotic linear Regge trajectory scales as $t^{1\over4}$ or $t^{3\over4}$. We then analyze the integral equation numerically and conclude that the massive ends solution is unique.


\subsec{What Are The WIHS Theories?}

Theories of weakly coupled higher spin particles are a natural generalization of the concept of classical strings. In flat space, these theories have the following properties. First, scattering amplitudes in these theories are meromorphic functions of the Mandelstam invariants. In other words, they are characterized by tree-level amplitudes with simple poles being their only singularities. The poles occur when the exchanged particles go on-shell. Second, the amplitude is unitary. At tree-level, it means that the residues of the poles are given by sums of partial waves with positive coefficients.\foot{In four dimensions partial waves are the usual Legendre polynomials. In general dimensions these are Gegenbauer polynomials.} Third, we assume that the amplitude is causal and that at least one particle with spin greater than two is being exchanged. Finally, we do not consider theories with an accumulation point in the spectrum.\foot{In AdS an example of such a theory is the Vasiliev theory \VasilievEN, dual to the $O(N)$ model \KlebanovJA. In the large $N$ limit it has an accumulation point that persists even when the higher spin symmetry is broken. On the other hand, the usual string theory, for any $\alpha'<\infty$ does not have an accumulation point. The accumulation point is only present when $\alpha' = \infty$. In this sense, $\alpha' = \infty$ is disconnected from $\alpha' \gg 1$. The same comment applies to the adjoint gauge theories.} Namely, we assume that for any finite energy $E$ there is only a finite number of particles with mass $m<E$.  Theories of WIHS are, thus, fully characterized by their spectrum and the correspondent three-point couplings.\foot{In this regard they are similar to CFTs.} In \CaronHuotICG\ we have shown that the large spin limit of these is universal and agrees with the one of the free string theory. Let us shortly review the argument of \CaronHuotICG.

\subsec{Review of the Leading Order Result}


In the approach of \CaronHuotICG\ one bootstraps the amplitude in the asymptotic regime of large and positive $s$ and $t$, which corresponds to imaginary scattering angles. In this kinematical regime, all contributions to the imaginary part of the amplitude are positive, with no possible cancelations. 
As a result, the amplitude is large (and is dominated by its imaginary part). This property is imposed on us by unitarity and comprises the dual of the Polchinski-Strassler mechanism from the amplitude bootstrap point of view. 

It turns out that in this regime, all WIHS amplitudes take the same universal stringy form \limitVen. Let us explain the main steps that lead one to this conclusion. Only the crude intuitive arguments will be presented here, and we refer the reader to \CaronHuotICG\ for further details. We also use this as an opportunity to spell out the notations. 

We consider the logarithm of the amplitude. At large $s$, every pole in $s$ leads to a logarithm with minus sign $-1\times\log(s)$. On the other hand, every zero in $s$ gives $+1\times\log(s)$. The Regge trajectory $j(t)$ can be thought of as a regulated sum of these zeros minus poles contributions to the coefficient of $\log s$. The finiteness of the trajectory tells us that there are finitely more zeros than poles (recall that $j(t)>0$ in the regime $t\gg 1$). 
Moreover, unitarity tells us that there is at least one zero between any two successive poles. Hence, beyond a certain value of $s$, there is exactly one zero between any two consecutive poles. This property allows us to express the amplitude in a simple Weierstrass product form 
\eqn\productformula{
A(s,t) = F(t) \prod_{i} \left( {1 - {s \over z_i(t)} \over 1 - {s \over m_i^2}} \right) \ .
}

While the spectrum of the theory, specified by $m_i$'s, is fixed, the location of zeros of the amplitude, specified by $z_i(t)$'s, depends on $t$. For any given $t>0$ we have unitarity zeros and excess zeros. Unitarity zeros are zeros that we can pair with the neighboring pole. The rest are excess zeros, we denote them by $z_i^e(t)$. There are approximately $j(t)$ of them. At large $t$ there are lots of them. While poles and unitarity zeros screen each other, the effect of the excess zeros adds up and dominate the amplitude at large $t$. This allows us to write an effective description of the logarithm of the amplitude in this regime in terms of a distribution, $\rho$, of the excess zeros in the complex $s$ plane
 \eqn\logAmplLead{\eqalign{
 \log A(s,t) &= \log F(t) + \int d^2 z \ \rho(t,z, \bar z) \left[\log \left( z - s \right)-\log(z)\right] + ... \ , \cr
 \rho(t,z, \bar z) &= \sum_{i} \delta^{(2)}(z - z_{i}^e(t)  ) \geq 0 \ .
 }}

Importantly, this representation of the amplitude is only valid in a small wedge around the positive $s$ axis, $A(s(1 + i \eps), t(1+i \eps))$. After we analytically continue \logAmplLead, it may not agree with the amplitude anymore. We have already encountered such behavior in the holographic setup. Due to the Polchinski-Strassler effect, the amplitude for real scattering angles generically differs drastically from the analytic continuation of its value for imaginary scattering angles.

The next important observation is that this is not an arbitrary distribution. Instead, to leading order we may think of $\rho$ as coming from a sum of partial waves with positive coefficients. Namely, we have that 
\eqn\partialwave{
\log A(s,t) \simeq \log  \int\limits_0^{j(t)} d j \  c_{j}(t) \ P_{j} \left(1 + {2 s \over t}\right)\ ,\qquad c_j(t)\ge0\ ,
}
where we have replaced the sum over spins at large $t$ by an integral. We observed that any such distributions that is consistent with the Regge behavior of the amplitude is confined within the unit circle $|{1 \over 2} + {z \over t}| \leq {1 \over 2}$.

The central result of \CaronHuotICG\ is that this picture is only consistent with crossing, unitarity and analyticity if the asymptotic trajectory is linear, $j(t)=\alpha't+\dots$ and the asymptotic density takes a simple form
\eqn\asymptoticrho{
\rho(x)=t\times\rho(t,tx,tx)=\left\{\eqalign{1&\quad-1<x<0 \ , \cr 0&\quad{\rm otherwise} \ .}\right. 
}
Correspondingly, the leading asymptotic of the amplitude is given by
\eqn\venezasym{\eqalign{
\log A(s,t) &= \alpha' t \int\limits_0^1 d x\rho(x)\log \left( 1 + {s \over t x} \right) + ... \cr
&=\alpha' \left[  (s+t) \log (s+t) - s \log s - t \log t \right] + ... \ .
}}

Notice that $\alpha'$ in the formula above is an arbitrary scale that is fixed in terms of the microscopic theory and is an arbitrary parameter from the bootstrap point of view.

Note that here we have chosen to place all the zeros inside the unit disk. One can, of course, charge the support of the distribution inside the unit disk without affecting the amplitude outside. The actual way the zeros are distributed only affects the scattering at real angles and will not be relevant for our analysis.

\subsec{Emergent Symmetry}

Before proceeding we note that even though we only imposed symmetry between the $s$ and the $t$ channels, the asymptotic density of excess zeros $\rho(x)$ exhibits a symmetry between the $s$ and the $u$ channels. Namely, it satisfies the relation
\eqn\symmetry{
\rho(x) = \rho(1-x)\ ,\qquad x=-{s\over t}\ . 
}
Indeed, $s \to u = -s -t$ transforms $x \to 1-x$. This is an emergent symmetry. In particular, \symmetry\ is not a microscopic symmetry of the Veneziano amplitude. It emerges in the asymptotic limit of the Veneziano amplitude because the microscopic zeros are equidistant. This symmetry of the microscopic distribution is a consequence of \ddis.

This symmetry is manifest for the scattering of identical particles in which case we only have even Legendre polynomials and the symmetry \symmetry\ is present at the microscopical level. The same argument applies to the sum of partial waves with odd spins only. It is, however, not a symmetry of a sum of both odd and even partial waves. Such sums appear in the scattering of non-identical particles.

\subsec{First Moment Constraint} 

Below we focus on the possible corrections to the asymptotic result \asymptoticrho. It turns out that one of the constraints on the asymptotic distribution that went into the derivation of \asymptoticrho\ also applies to the corrections. This is a constraint on the first moment of the distribution
\eqn\firstmomentconstraint{
M_1=-\int d^2z\,\rho(z,\bar z)\,z\ge{1\over2} \ .
}
Since the distribution \asymptoticrho\ saturates this constraint, it just go through to the correction. Namely,
\eqn\firstmomentconstraintdelta{
\delta M_1=-\int d^2z\,\delta\rho(z,\bar z)\,z\ge 0 \ .
}

As we will see in the next section, the massive endpoints correction saturates \firstmomentconstraintdelta.

\subsec{Lifting the Degeneracy}

The leading asymptotic form of the amplitude \venezasym\ is insensitive to whether the spectrum is degenerate or not. The reason is that when we evaluate the amplitude at $A(s(1+i\epsilon),t(1+i\epsilon))$ for large $s$ and $t$, we are far from the real axis compared to the separation between adjacent poles. Far from the real axis, nearby poles look like as if they were degenerate, see \CaronHuotICG\ for more details. As we decrease $|s|$ and $|t|$, we have to start taking into account corrections to this approximation.

The Veneziano amplitude corresponds to a maximally degenerate spectrum. More precisely, all resonances on subleading trajectories are exactly degenerate with the leading one. This situation is very non-generic and is due to a large symmetry of free strings \refs{\GrossUE,\MooreNS}. In the absence of such a symmetry, the degeneracy is lifted. For example, in theories with an effective long string description (such as large $N$ QCD), this can be seen explicitly by studying the spectrum in the effective theory \AharonyDB.

Suppose we are in a generic situation where the spectrum is not degenerate. How would it affect the asymptotic distribution of excess zeros discussed above? To answer this question let us consider a pole, $m_i$, on a sub-leading Regge trajectory $j_{sub-leading}(t)<j_{leading}(t)$ and let us assume it is not degenerate with the leading Regge trajectory. At $t\sim m_i^2$ the asymptotic distribution is controlled by $j_{leading}(m_i^2)$ excess zeros. However, at $t=m_i^2$ the amplitude is a polynomial in $s$ of integer degree $j_{sub-leading}(m_i^2)<j_{leading}(m_i^2)$. Hence, as we move in the $t$ complex plane close to $t=m_i^2$, $j_{leading}(m_i^2)-j_{sub-leading}(m_i^2)$ excess zeros must escape to infinity. Similar jumps occur whenever we move between adjacent poles on different trajectories. By the time we reach a pole on the leading trajectory, the number of excess zeros is back to $j_{leading}(t)$. 
\ifig\general{The behavior of excess zeros $z^{(e)}_i(t)$ as we change $t$. As we cross the poles of the sub-leading Regge trajectories, some of the zeros escape to infinity and come back to the unit disk. This effect due to non-degeneracy of the spectrum is responsible for the non-zero support of the excess zeros distribution outside of the unit circle. } {\epsfxsize5.2in\epsfbox{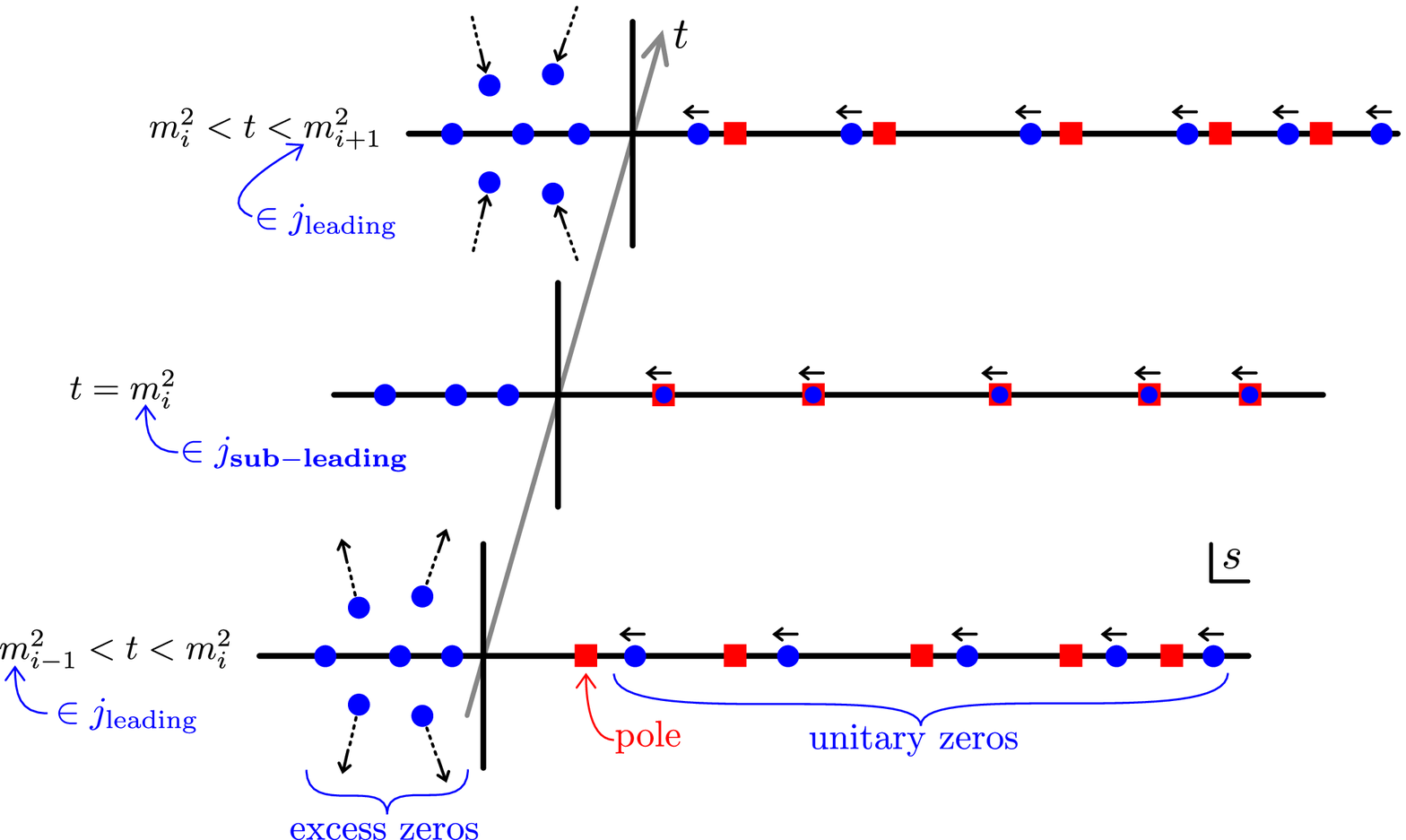}}

When we consider $A(s(1+i\epsilon),t(1+i\epsilon))$ and look at the next-to-leading correction in $t$ (or $s$), we start seeing the effect of these jumps. Since we are still far from the real axis, what we observe is an average effect of these jumps. The correction to the asymptotic distribution of excess zeros, thus, could have a contribution outside of the unit circle. In addition, we have a small disruption of the constant distribution \asymptoticrho\ inside the unit circle. While the latter could be both positive and negative, the former must be positive. 

Importantly, any other type of correction that is insensitive to the spectrum degeneracy cannot lead to a distribution of excess zeros outside of the unit circle \CaronHuotICG. We conclude that any correction to the asymptotic distribution outside of the unit circle results from the fact that the spectrum is non-degenerate. Consistent with that picture, for the Veneziano amplitude there is exact degeneracy and no correction to the asymptotic linear trajectory or distribution.

As in \CaronHuotICG, if we assume separation of scales we get
\eqn\corroutside{
\delta\log A(s,t)=m^{2-2 k} t^k\int d^2z\,\delta\rho(z,\bar z)\log\left(1+{s\over t z}\right) \ ,
}
where $\delta\rho(z,\bar z)$ is real and has positive support outside of the unit circle. 
After putting back the dimensionfull parameters, the corrected Regge trajectory is
\eqn\trajectoryk{
j(t)=\alpha'\left(t+c\,m^{2-2k}t^k+\dots\right) , \qquad{\rm where}\qquad\int d^2z\,\delta\rho(z,\bar z)=c \ .
}
Here, $m$ has mass dimensions and $c$ is dimensionless constant. 

Recall that the leading distribution \asymptoticrho\ has an emergent symmetry between the $s$- and the $u$-channels, \symmetry . As discussed above, this is also a symmetry of zeros of any individual partial wave. Once we consider the correction due to the non-degenerate spectrum, that symmetry could be broken. However, such breaking requires fine-tuning. To break the symmetry, the non-degenerate spectrum should have two scales. At the first scale, only part of the degeneracy is removed by grouping together even and odd partial waves so that the zeros of their sums do not respect the $z\leftrightarrow1-z$ symmetry at the macroscopic level. As a result, in between, the excess zeros escape to infinity in a non-symmetric way. At the second scale, all the degeneracy is removed. 

In a generic situation, we expect to have one scale where all the degeneracy is lifted. The corresponding pattern of excess zeros outside of the unit circle then should respect the symmetry to the $u$ channel. Namely, in a generic situation, we expect that 
\eqn\symmetrycorr{
\delta\rho(1-z,1-\bar z)=\delta\rho(z,\bar z)=\delta\rho(\bar z,z)\ ,
}
where the last equality follows from the reality of the amplitude.

\subsec{Non-generic Degenerate Correction}

Before proceeding to the massive endpoints correction, let us discuss shortly a non-generic scenario in which the spectrum is degenerate, and the symmetry \symmetrycorr\ is broken. In such a scenario there are of course no excess zeros outside of the unit circle. The only effect of the correction is to re-distribute zeros inside the unit circle. Microscopically, the amplitude is still controlled by the sum of Legendre polynomials, as in \partialwave, but with a slightly corrected coefficients and spectrum.

Given $k$ in \trajectoryk, analyticity and crossing fix the solution uniquely, \CaronHuotICG. The corresponding correction to the density of excess zeros has support in the interval $0 < x < 1$ and is given by
\eqn\uniqueresult{
\delta \rho_k (x)= t^{2-k} \times \delta \rho(t,tx,tx)= {k \sin \pi k \over \pi} x^{k-1} \left( - \log x + \sum_{m=1}^{\infty} (k)_m {1 - x^m \over m \ m!}\right)
}

Importantly, for any $k \neq 1$ we have $\delta \rho_k (x) \neq \delta \rho_k(1-x)$, so the symmetry to the $u$-channel is broken.

On the other hand, for scattering of identical particles, the correction \uniqueresult\ is ruled out even if the spectrum is degenerate. The reason is that in the asymptotic limit, \partialwave\ contains only even partial waves. As a result, the symmetry to the $u$-channel is explicit. Hence, a Regge trajectory that consists of degenerate spectrum of even spins only takes the form
\eqn\lrdeg{
j(t) = \alpha' t  + \alpha_0 +\dots\ ,
}
where the dots stand for terms that are suppressed at large $t$.

\subsec{Massive Ends and Excess Zeros}

In this section, we analyze the massive endpoints correction \differentmasses\ using bootstrap. We find that it indeed satisfies all constraints that were discussed above and, hence,  is captured by the dynamics of excess zeros. We then discuss additional properties of the result and the uniqueness of the amplitude of this type.  

Because the terms in \differentmasses\ are independent, we will focus on one of them only. Namely, we set $m_{2,3}=m_{4,1}=0$ and $m_{1,2}=m_{3,4}=m$ in \differentmasses
\eqn\correctiononemass{
\delta\log A(s,t)=-{16 \sqrt{\pi} \alpha' \over 3}m^{3/2}\left( {s\, t \over s + t} \right)^{1/4} K\left({s \over s + t} \right)
}
This amplitude is of course no longer crossing symmetric. When talking about the correction, crossing is not an important symmetry. It can always be restored by summing over the two independent corrections due to the mass in the $s$ and the $t$ channels, \mainagain.

Now, \correctiononemass\ represents an amplitude where the string that is been exchanged in the $t$-channel (that is giving $s^{j(t)}$) has massive endpoints while the ends of the string that is been exchanged in the $s$-channel are still massless. As a result, the correction to the trajectories in the two channels takes the form
\eqn\trajectorycorr{
\delta j_s(s) = 0\ ,\qquad\delta j_t(t) =- t^{1/4} \ ,
}
where for convenience, we work in units where ${8 \sqrt{\pi} \alpha' \over 3}m^{3/2} = 1$. It is easy to verify that 
\eqn\trajectoryst{
\lim_{t \to \infty}\delta\log A(s,t)= - \pi\, s^{{1 \over 4}} +\dots\qquad{\rm and}\qquad\lim_{s \to \infty}\delta\log A(s,t)= -\, t^{{1 \over 4}}\log s +\dots
}

Because the trajectory in the $s$-channel is not corrected, the corresponding correction to the distribution of excess zeros should integrate to zero. Importantly, the degeneracy in the $s$-channel is not lifted. Therefore, this correction to the distribution is expected to be bounded to the unit disk. Indeed, we find that
\eqn\correctiononemass{
- 2 \left( {s\, t \over s + t} \right)^{1\over4} K\left({s \over s + t} \right)= s^{{1 \over 4}} \int\limits_0^1 d x \ \rho_s (x) \log (1 + {t \over s x})
}
where
\eqn\representation{
\rho_s(x) = r(x)+r(1-x)\ ,\qquad r(x) = - {1 \over 2 \sqrt 2 \pi} {K(x) - 2 E(x) \over [x(1-x)]^{3/4}}  \ ,
}
and
\eqn\ellipticKandE{\eqalign{
K(x) &= \int\limits_0^{1} {d t\over \sqrt{1 - t^2} \sqrt{1 - x t^2}}\ ,\qquad E(x) = \int\limits_0^{1} d t \ { \sqrt{1 - x t^2} \over \sqrt{1 - t^2}} \ ,
}}
 are complete elliptic integrals of the first and second kind correspondingly. One can check that, indeed,
\eqn\integratetozero{
\int\limits_0^1dx\,\rho_s(x)=0 \ .
}

The correction to the distribution in the $t$-channel is more interesting. It takes the following form
\ifig\general{The correction to the distribution in the $t$-channel has two supports. One is an horizontal distribution, $\rho_{t,h}$. It is bounded inside the unit circle and is of indefinite sign. The other is a vertical distribution, $\rho_{t,v}$. It is a strictly positive distributed that is diffused along the line $\beta=s/t=-1/2+iy$ and goes all the way to infinity.} {\epsfxsize1.8in\epsfbox{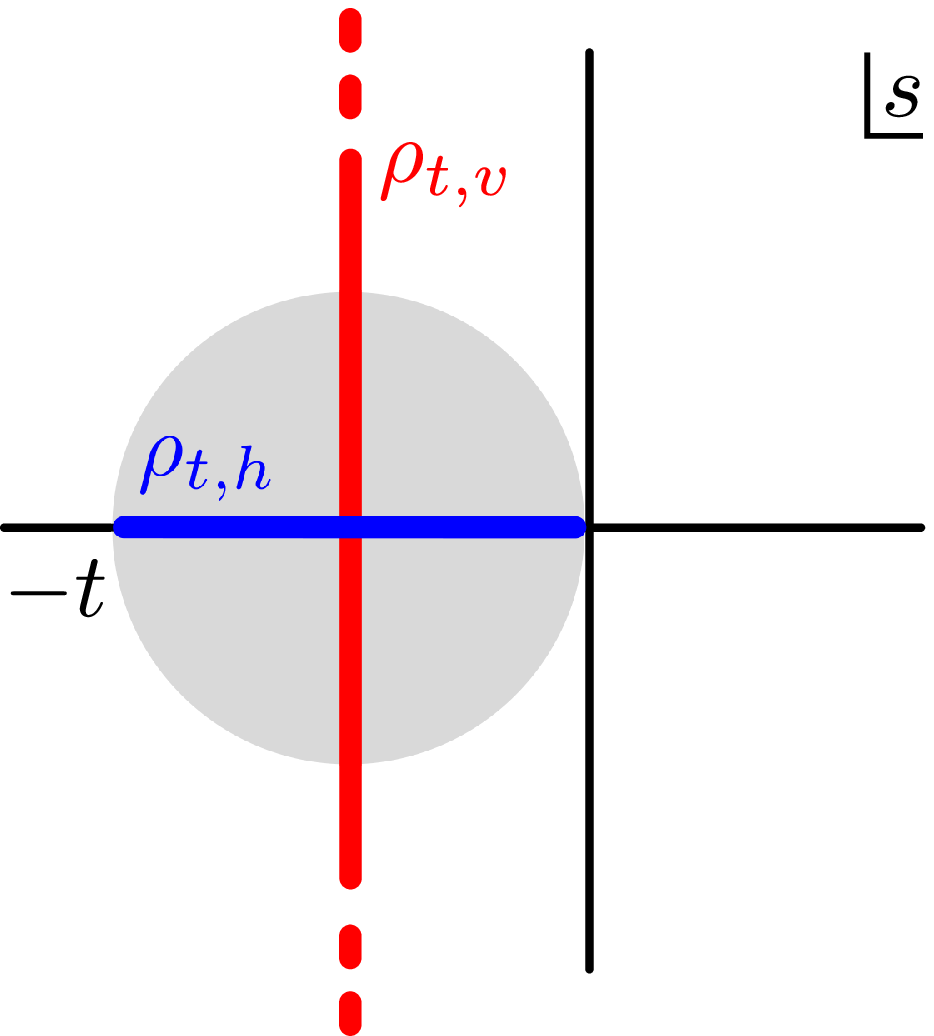}}
\eqn\schannel{
- 2 \left( {s t \over s + t} \right)^{1\over 4}\!\! K\left({s \over s + t} \right) = t^{{1 \over 4}}\!\!\int\limits_0^1\! dx \ \rho_{t,h}(x) \log (1+ {s \over t x})+ t^{{1 \over 4}}\!\!  \int\limits_{-\infty}^\infty\!\! dy \ \rho_{t,v}(y) \log ( 1 + {s \over t (1/2+i y)}  ).
}
where the horizontal (first) term is given by a symmetric $\rho_{t,h}(x)= \rho_{t,h}(1-x)$ distribution of zeros between $(-1,0)$
\eqn\rhoh{\rho_{t,h}(x)=\left\{\eqalign{&-r(x)\qquad\quad{\rm for}\quad 0<x<{1 \over 2} \ ,\cr 
&-r(1-x)\quad\ {\rm for}\quad {1 \over 2}<x<1 \ .} \right.}
The vertical (second) piece, running along $x={1\over2}+iy$, is given by 
\eqn\rhov{
\rho_{t,v}(y) = {4\over \pi} z^{1/2} (1-z)^{1/2} \left( E(z) - (1-z) K(z) \right), \qquad{\rm where}\qquad
z = {1 \over 2} -\sqrt {y^2 \over 1 + 4 y^2} \ .
} 

Again to correctly reproduce the Regge limit $s \gg 1$, namely $- t^{1/4} \log s$, we have
\eqn\identity{
\int\limits_0^1 d x \ \rho_{t, h} (x) + \int\limits_{- \infty}^{\infty} d y \ \rho_{t,v}(y) = - 1 .
}
Moreover, as one can easily check $\rho_{t,v}(y) > 0$ for any $y$ and the symmetry $x \to 1- x$ is manifest. Finally, we note that this distribution saturate the condition on its first moment. Namely, $\delta M_1=0$ in \firstmomentconstraintdelta.

Recall that from the holographic and EFT points of view the sign of the correction to the Regge trajectory $\delta j(t) = c t^{1/4}$ is fixed because it corresponds to the slowdown of the string, namely $c<0$. From the bootstrap point of view, this condition follows neatly from positivity of $\rho_{t,v}(y)$. 

To summarize, we see that the massive endpoints correction is well-described by the effective classical distribution of excess zeros. It satisfies all the known bootstrap restrictions. Since it has support outside of the unit disk, it results from a non-degenerate spectrum. 

\subsec{Discontinuities of the Massive Endpoints Correction}

The correction $\delta\log A(s,t)$ in \correctiononemass\ has some remarkable properties that we now describe. As discussed above, it corresponds to a situation where in the $t$-channel the Regge trajectory is corrected and in the $s$-channel it is not. Correspondingly, in the $t$-channel we have excess zeros outside of the unit disk and in the $s$-channel we do not. 

It is instructive to introduce the effective electric field in both channels and express them in terms of the distribution in the $s$-channel
\eqn\genfunczB{\eqalign{
f_s (\beta) &\equiv s^{1-k} \pa_t \delta \log A(s,t) = \int\limits_0^{1} d x {\rho_s(x) \over x + {1 \over \beta}} \ , ~~~ \beta = {s \over t} \  , \cr
f_t (\beta) &\equiv t^{1-k} \pa_s \delta \log A(s,t) = \beta^{k-1}\left[k\int\limits_0^1 dx\,\rho_s(x)\log(1+{1\over\beta x})-\int\limits_0^1 dx{\rho_s(x)\over1+\beta x}\right] \ ,
}}
where for massive endpoints, we have $k={1 \over 4}$ and $\rho_s(x)$ is given in \representation .

It is easy to check that for these specific $k$ and $\rho_s(x)$, the electric field \genfunczB\ have an intriguing property
\eqn\crossingfirst{
{1 \over 2 i} \left[ f_t(-\beta+ i 0) - f_t(-\beta - i 0)  \right] = f_s (- {1 \over \beta})\ ,\qquad \beta > 1  \ .
}
One way to interpret it is the following. The LHS of \crossingfirst\ effectively computes the vertical density of zeros $\rho_{t,v}(y)$, whereas the RHS relates it to the electric field created by the zeros inside the unit circle $\rho_s(x)$ in the dual channel. 

Another property is a direct consequence of \ddis\ which after differentiation takes the form\foot{It could be also derived from \crossingfirst\ using $ f_t({-1\over\beta+ i 0}) + f_t({-1\over\beta - i 0})=0$ together with $\rho_t(z) = \rho_t(1-z)$.}
\eqn\doubledisc{
{\rm dDisc}[f_t(\beta)] = f_t(\beta) + {1 \over 2} \left[ f_t(-1-\beta+ i 0) + f_t(-1-\beta - i 0)  \right] = 0\ ,\qquad\beta > 1 \ .
}
This property of the electric field also holds for the logarithm of the amplitude. As we explained this is a condition expected for any worldsheet theory. In terms of the bootstrap it is related to the $s - u$ channel crossing and the statement that there is no Stokes phenomenon when we connect two regions of scattering at purely imaginary angles in each of which the amplitude is exponentially large.

Below we will argue that \doubledisc\ fixes the solution uniquely to be given by \representation\ and $k=1/4$. To prepare the ground, we will first analyze the large $\beta$ structure of $f_t(\beta)$ for general $k$.

The double discontinuity constraint is simpler in terms of the symmetric variable
\eqn\newvariable{
b = \beta - {1 \over 2} .
}
In terms of this variable, the analytic continuations in \doubledisc\ simply rotates $b \to -b$. The general large $b$ power expansion of the electric field that is consisting with zero double discontinuity is of the form
\eqn\consistent{
\lim_{b\to\infty} f_t(b) = \sum_{n}^{\infty} {1 \over b^{1+2 n}} \left( c_n + d_n \log b \right) \ .
}

Lets now see what this implies for the small $x$ expansion of $\rho_s(x)$. Using the symmetry $\rho_s(x) = \rho_s(1-x)$ we rewrite $f_t (\beta)$ in \genfunczB\ as
\eqn\rewrite{
f_t (\beta) = \beta^{k-1} \int\limits_0^{{1 \over 2}} d x  \left[ F(\beta,x) + F(\beta,1-x) \right]\rho_s(x) \ ,\qquad
F(\beta,x) = k \ \log(1+{1\over\beta x}) - {1 \over1+\beta x} \ .
}
From \rewrite\ we see that the large $b$ expansion of $f_t(1/2+b)$ maps into the small $x\to0$ limit of $\rho_s(x)$ or its mirror point $x\to1$. Since, when writing \rewrite\ we are already imposing the $\rho_s(x)=\rho_s(1-x)$ reflection symmetry, its enough to focus on the small $x$ expanssion.

We start by recalling that from the Regge limit we already know that 
\eqn\regge{
\lim_{x\to 0}\rho_s(x) \sim x^{k-1} \log x + [{\rm less \ singular\ terms}] \ .
}
Plugging $x^{k-1} \log x$ in \rewrite\ and expanding at large $b$ we get both, terms of the type $b^{k-1-n}$ and of the type $b^{-n}$, with $n$ integer. 

We conclude that the sub-leading terms in the small $x$ expansion \regge\ should be such that
\item{a)} They should produce terms $d_n b^{-2n-1}\log b$ in \consistent\ with some $d_n \neq 0$. These are the terms that correspond to a distribution of excess zeros outside the unit disk and represents the removal of the degeneracy.\foot{If the degeneracy is not removed then as was shown in REF there is no solution to the problem that we are considering here.}
\item{b)} It has to by such that all terms of the type $b^{-2n} \log b$ must cancel.
\item{c)} It has to by such that all terms of the type $b^{-2n}$ must cancel.
\item{d)} The terms of the type $b^{k-1-n}$ that are generically produced from \rewrite\ should all be canceled.

These conditions are satisfied as follows
\item{A)} To generate the terms terms $d_n b^{-2n-1}\log(g)$ with $d_n \neq 0$ we add to $\rho_s(x)$ sub-leading terms of the type $b_n x^{k-1+n} \log x$.
\item{B)} To cancel $d_n$ with even $n$ we tune the coefficients $b_n$. One can easily check that it fixes $b_{2k+1}$ in terms of $b_{2k}$.
\item{C)} Next we can ask: which terms can we add to the distribution so that it generates the same terms that we already have? The only other terms that have this property are terms $a_n x^{k-1+n}$. These generate terms $b^{k-1-n}$ and $b^{-n}$ and, thus, could be used to satisfy c) and d). 

\

Finally, one can try and add terms of the type $x^n$ to the small $x$ expansion of $\rho_s(x)$. However, it is easy to see that these terms generate new undesired terms of the type $b^{k-1-n} \log b$, which should be all set to zero. This condition sets the coefficients of such $x^n$ terms to zero. 

Thus, we arrive at the following ansatz for the distribution
\eqn\smallyexpansion{
\lim_{x\to0}\rho_s (x) = x^{k-1} \sum_{n=0}^{\infty} x^n (a_n \log x + b_n) \ .
}
Next, we explore this expansion in greater detail.

\subsec{Fixing The Solution}

We shall now utilize the $s-u$ crossing \ddis\ in the form of \doubledisc\ to fix $k$ and the distribution. That is, we consider a flavor amplitude with a generic $k$ and try to find $\rho_k(x) = \rho_k(1-x)$ such that \doubledisc\ holds. This constraint leads us to the massive ends amplitude as the unique solution.

To start, we use \genfunczB\ to turn the zero double discontinuity condition \doubledisc\ into an integral equation. More precisely, we can compute $f_t(-1-\beta \pm i 0)$ by doing an analytic continuation under the integral \genfunczB . As we do that ${1 \over 1 + x \beta}$ and $\log(1+{1\over\beta x})$ develop an imaginary part. This imaginary part, however, cancels in the combination \doubledisc . As the result we get an integral equation of the type $\int_0^1 d x \ \tilde K(\beta, x) \rho_k (x) = 0$ for $\beta>1$. Introducing $y = {1 \over \beta}$ and taking the combination of the equation together with its image under $y \to 1-y$ and $x \to 1-x$, we turn it into a simpler integral equation. Using the symmetry $\rho_k (x)  = \rho_{k} (1-x)$, we can write the result as follows
\eqn\integralequation{
\delta \rho_k (y) = \int\limits_0^1 d x \,\left[ K(y,x) + K(1-y,1-x)\right] \delta \rho_k (x) \ ,
}
where, after some algebra, the kernel can be written as\foot{Here, ${\rm P}{1 \over x-y}= \lim_{\eps \to 0} {1 \over 2} \left( {1 \over x-y + i \eps} + {1 \over x-y - i \eps}  \right)$ stands for the principle value.}
\eqn\kernel{\eqalign{
K(y,x) &= {\cot \pi k \over \pi} \left[ y\,{\rm P }{1 \over x-y}  - k \log {x \over |x-y|} \right] \cr
&+ {(1-y)^{k-1} \over \pi \sin \pi k} \left[ {y \over x+y- x\, y} + k \log {x\, (1-y) \over x+ y- x\,y } \right]  \ .
}}

It is straightforward to check that the distribution \representation\ indeed solves this integral equation. To proceed we rewrite the integral equation \integralequation\ as follows
\eqn\rewriteB{
 \rho_k (y) = \int\limits_0^{1\over 2} d x \,\left[ K(y,x) + K(1-y,x) + K(x,1-y) + K(1-y,1-x)\right]  \rho_k (x) ,
}
where we again used the symmetry $ \rho_k (x)  =  \rho_k (1-x)$. At this stage we can plug the ansatz \smallyexpansion\ to both sides of the equation and match the coefficients. 

We focus on the terms $a_n\,y^{k+n-1} \log y$ in \integralequation. For the first few $n$'s one finds the following result
\eqn\consistencycheck{
a_1 =0\ ,\qquad 1 - \cot^2 \pi k = 0 \ .
}
The second equation has two solutions
\eqn\solutionsfork{
k = {1 \over 4}\qquad{\rm and}\qquad k = {3 \over 4} \ .
}
Higher order orders terms in $n$ lead to an infinite number of relations between different $a_n$'s, again fixing $a_{2 n+1}$ in terms of $a_{2n}$.

After this point we have not found an analytic way to solve the equation \rewriteB . Its structure, however, suggests a straightforward numerical strategy, as we now describe. We truncate the sum \integralequation\ at some $n_{max}$
\eqn\numerics{
 \rho_k^{num} (x) = x^{k-1} \sum_{n=0}^{n_{max}} x^n (a_n \log x + b_n) \ .
}
We then consider the positive-definite functional by squaring the difference between the correspondent coefficients in the LHS and RHS of \rewriteB 
\eqn\functional{
\chi^2 = \sum_{n=0}^{n_{max}} (a_n^{LHS} - a_n^{RHS})^2 + (b_n^{LHS} - b_n^{RHS})^2
}
and minimize it for arbitrary values of $a_{n>1}$ and $b_n$. If a solution exists then we expect that by increasing $n_{max}$ we will get $\chi^2$ approaching zero. This is, indeed, exactly what we got for $k =1/4$. It is however not the case for $k =3/4$. Hence, we conclude that only solutions with $k=1/4$ exists. 

To probe the uniqueness of the $k =1/4$ solution we subtract the massive end solution from the other potential solution $\delta \hat \rho_k$, so that $a_0 = 0$. Since the integral equation  \rewriteB\ is linear, the subtracted solution $\delta \hat \rho - \delta \rho^{\rm massive\ ends}$ is again a solution. We now assume that there is such $\hat a_{n} - a_n^{massive \ ends} \neq 0$ and repeat the minimization procedure for this conditions. We checked for few low $n_{max}$'s that $\chi^2 \simeq 0$ solution does not exist.  We conclude that $\hat a_n = a_n$ for all $n$. Repeating the same procedure for $b_n$'s we conclude that the massive ends solution is unique.

Let us emphasize that given a $\delta \rho_k(x)$ that satisfies \integralequation\ we can easily use it to construct a fully crossing symmetric amplitude by simply taking a sum of 
\eqn\corrCC{
\delta \log A(s,t) =\int\limits_0^1 d x \  \rho_k(x) \left( t^k \log (1 + {s \over t\, x})  + s^k \log (1 + {t \over s\, x}) \right) \ .
} 
This amplitude could be still rewritten in terms of zeros in one channel only and moreover by construction it has zero double discontinuity \ddis. 

\subsec{The Role of Crossing}

Note that in the discussion above crossing did not play any role. Indeed, motivated by the picture of the string we assumed: 

\item{1.} The correction to the amplitude is a sum of $t$ and $s$ channels corrections, \corrCC .

\item{2.} The double discontinuity \ddis\ of the correction vanishes.

Given these two assumptions, we showed that the massive ends correction is unique. Both of these assumptions are tied to the fact that the massive endpoint correction is a boundary interaction on the string worldsheet. The first one comes about because in perturbation theory on the worldsheet we are instructed to sum over the boundary interaction in the $s$ and $t$ channels. The second one comes about because the bulk of the string is free, (and hence, holds at any order in the boundary interaction). 

In general, we would like to relax both assumptions and still show that the solution is unique. Since both are associated with a boundary interaction on the worldsheet, they may be not independent of each other. From our discussion of the structure of the degeneracy removal, it seems that assumption 1 is not enough to fix the solution. Instead, we expect assumption 1 to follow from assumption 2.

Dropping assumption 1, we can proceed as follows. Instead of trivializing crossing and analyzing the double discontinuity constraint as we did above, we could trivialize the latter and let the non-trivial constraint to come from crossing. More precisely, we imagine the following problem. Consider a crossing symmetric correction to the amplitude $\delta \log A(s,t) = \delta \log A(t,s)$ with zero double discontinuity \ddis. Based on our discussion we expect this correction to take the form
\eqn\correction{
 \delta \log A(s,t) = t^{k}\!\!\int\limits_0^1\! dx \ \rho_{h}(x) \log (1+ {s \over t x})+ t^{k}\!\!  \int\limits_{-\infty}^\infty\!\! dy \ \rho_{v}(y) \log ( 1 + {s \over t (1/2+i y)})\ .
}
We can now first impose the vanishing of the double discontinuity. The double discontinuity outside of the unit disk takes the form
\eqn\doublediscsimple{
\rho_{v}( y e^{i {\pi \over 2}} ) + \rho_{v}(-y e^{-i {\pi \over 2}} ) = 0 \ , ~~~ y > {1 \over 2} \ .
}
This condition fixes the form of the large $y$ expansion of $\rho_{v}(y)$ to be
\eqn\writesolution{
\rho_{v}(y) = \sum_{n}{c_n \over |y|^{2 n+1} } \ .
}

Now the crossing symmetry constraint becomes very nontrivial. Indeed, we have
\eqn\crossing{\eqalign{
 &t^{k}\!\!\int\limits_0^1\! dx \ \rho_{h}(x) \log (1+ {s \over t x})+ t^{k}\!\!  \int\limits_{-\infty}^\infty\!\! dy \ \rho_{v}(y) \log ( 1 + {s \over t ({1\over2}+i y)}  ) \cr
 &= s^{k}\!\!\int\limits_0^1\! dx \ \rho_{h}(x) \log (1+ {t \over s x})+ s^{k}\!\!  \int\limits_{-\infty}^\infty\!\! dy \ \rho_{v}(y) \log ( 1 + {t \over s ({1\over2}+i y)})\ ,
}}
where $\rho_{v}(y)$ is subject to \writesolution .

Showing that the massive ends solution is the unique solution to this problem is equivalent to relaxing assumption 1 above. We expect this to be the case, but we leave the analysis of this problem to the future.

\newsec{Conclusions}

In this paper, we considered the $2\to2$ scattering amplitude in theories of weakly interacting higher spin particles. We have computed the first correction to the asymptotic form of the amplitude \limitVen. The result is \mainresult.  We have argued that this is the unique universal correction in the large energy, imaginary scattering angle regime. In the string theory description, this correction due to massive endpoints that slow down string's motion. On the bootstrap side, \mainresult\ captures the non-degeneracy of the microscopic spectrum, which the leading solution \limitVen\ does not reflect. Similarly, the correction to the $n$-point amplitude takes the form \onshellaction . In principle, one can compute further corrections which are sub-leading in $m$, but we do not expect these to be universal. It is also trivial to generalize the result to the case of scattering of mesons with different flavors \differentmassesn. 

Both the leading solution \limitVen\ and the correction \mainresult\ exhibit an emergent $s \leftrightarrow u$ channel symmetry. We explained how it comes about from the analytic continuation of a free string classical solution with boundary interactions. The continuation develops an imaginary part which computes the on-shell action on the Lorentzian continuation of the Euclidean solution. The real part of the on-shell action, however, stays the same as a result of a certain stringy Stokes-like phenomenon. The consequence of this is the fact that the double discontinuity vanishes \ddis.

We have analyzed the result using the bootstrap techniques of \CaronHuotICG . The new feature of the correction is non-boundedness of the support of the excess zeros. This property reflects the sensitivity of the asymptotic amplitude to the non-degeneracy of the microscopic spectrum (the leading order answer is insensitive to it). 
From the bootstrap point of view, the property \ddis\ could be interpreted as absence of the Stokes phenomenon when we connect different channels that describe scattering of identical particles at imaginary angles. It would be interesting to understand this symmetry better and see if it could be violated.
When \ddis\ is imposed, we have shown that the massive endpoints solution is unique. 

We have argued that \mainresult\ is the only universal, model-independent correction to \limitVen\ which grows with energy. To bootstrap the amplitude beyond this correction we believe one should start adding model-dependant input. It would be interesting to understand what is the minimal input needed to fix the next correction and what is the most natural way to specify it. One might hope that akin to the 3d Ising model \ElShowkHT, there exist special corners in the space of parameters of the spectrum that describe physical theories. One natural assumption is to have graviton in the spectrum. More generally, it would be most rewarding if one could formulate the problem of fixing the amplitude using a numerical $S$-matrix bootstrap approach, (see \refs{\PaulosFAP,\PaulosBUT} for a different, but related recent progress). 

There is one important characteristic feature of strings that so far our analysis did not touch. That is, the high energy density of states in the string models exhibits characteristic Hagedorn growth. For the Veneziano amplitude, due to the degeneracy of the spectrum, this property could not be seen at the level of the four-point amplitude.\foot{ One can argue for it based on the four-point function if one assumes that all the three-point coefficients are of the same order. This extra assumption, however, needs further justification.} There, the Hagedorn growth was derived by considering higher-point functions \FubiniQB, \FubiniWP. We expect this to be a necessary step to establish Hagedorn from the bootstrap as well. We note, however, that the massive endpoints correction results from the removal of degeneracy in the spectrum. Hence, it may be tied to the Hagedorn growth. To understand if this is indeed the case one should study what is the minimal density of states that could lead to the correction \mainresult. We leave a more thorough investigation of the degeneracy removal pattern to the future.

Another natural direction is to consider full quantum amplitudes \refs{\PaulosFAP,\PaulosBUT}. In this case, akin to the Polchinski-Strassler mechanism, the large energy limit of the gravitational amplitudes seems to have universal features due to the production of large black holes. It would be fascinating to explore this regime using bootstrap.

Finally, let us mention an interesting extension of the present analysis. That is the analysis of WIHS theories in AdS. Such theories are dual to large $N$ CFTs. A natural space for their bootstrap is Mellin space \PenedonesUE . It is quite easy to recover the usual results about the gauge theories, see Appendix B. In AdS, however, we expect to have more solutions, corresponding to vector models, see e.g. \GiombiMS,  and SYK-like models, see e.g. \MuruganETO\ and references therein. These models have an accumulation point in the twist spectrum, which is an analog of the accumulation in the spectrum discussed in the flat space context. It would be very interesting to understand what are all possible WIHS in AdS and bootstrap them. We leave this problem for the future.

\newsec{Acknowledgments}

We are grateful to O. Aharony, S. Caron-Huot, T. Cohen, S. Dubovsky, R. Gopakumar, Z. Komargodski, S. Komatsu, C. Sonnenschein, M. Strassler and S. Yankielowicz for useful discussions. The work of AZ is supported by a Simons Investigator Award of X.~Yin from the Simons Foundation. AS has been supported by the I-CORE Program of the Planning and Budgeting Committee, The Israel Science Foundation (grant No. 1937/12) and by the Israel Science Foundation (grant number 968/15) and the EU-FP7 Marie Curie, CIG fellowship.

\newsec{Appendix A: Mass Perturbation for the Four-point Amplitude}

In this appendix, we develop the small mass perturbation theory, order by order. We confirm \cubic\ and \onshellaction\ through a detailed step-by-step computation. This exercise also demonstrates the power of the shortcut that we took in the main text.

Recall that the small mass perturbation theory is organized in powers of $\sqrt{m}$ (this fact was derived in section 3). Hence, we expand the action 
\eqn\actionagain{
S = {1 \over 2 \pi \alpha'} \int\! d^2 z\, \pa_z x^{\mu} \pa_{\bar z} x_{\mu} + {1 \over 2} \int d \sigma \left( e\,\pa_\sigma x^{\mu} \pa_{\sigma} x_{\mu} + {m^2 \over e}\right) + i \sum_j k_j^{\mu} x_{\mu} (\tilde \sigma_j) \ ,
}
as well as all variables in powers of $\sqrt{m}$. We then extremize the action order by order. We have
\eqn\expand{
S=S_0+\sqrt m S_1+\dots\ ,\qquad x^\mu=x_0^\mu+ \sqrt m x_1^\mu+\dots\ ,\qquad e=e_0+\sqrt m e_1+\dots 
}
We also have to expand the vertex operators insertion points which enter the on-shell action through an $SL(2)$ invariant cross ratio ${\tilde\sigma_{1,2}\tilde\sigma_{3,4}\over\tilde\sigma_{1,4}\tilde\sigma_{2,3}} $.

Indeed, the action \actionagain\ has an $SL(2)$ gauge symmetry that acts on the vertex operator insertion points $\tilde \sigma_j$, leaving their conformal cross ratio ${\tilde\sigma_{1,2}\tilde\sigma_{3,4}\over\tilde\sigma_{1,4}\tilde\sigma_{2,3}} $ invariant. It is convenient to pick an $SL(2)$ frame. We choose
\eqn\lambdagauge{\eqalign{
\{\check \sigma_1,\check \sigma_2,\check \sigma_3,\check \sigma_4\}=&\left\{0,{1\over2},{s+t\over s+2t},1\right\}
-\{ \sqrt m \lambda_1+ m \lambda_2+\dots,0,0,0\}\cr
\equiv&\ \ \{\sigma_1,\sigma_2,\sigma_3, \sigma_4\}\ \,-\{\sqrt m \lambda_1+ m \lambda_2+\dots,0,0,0\} \ ,
}}
where we have used the fact that ${\tilde\sigma_{1,2}\tilde\sigma_{3,4}\over\tilde\sigma_{1,4}\tilde\sigma_{2,3}}|_{m=0} \equiv \lambda_0 = s/t$, \refs{\GrossKZA,\GrossGE}. In terms of these $\lambda$'s the cross ratio takes the form
\eqn\lambdaexp{
{\tilde\sigma_{1,2}\tilde\sigma_{3,4}\over\tilde\sigma_{1,4}\tilde\sigma_{2,3}}=\lambda_0+ \sqrt m  \lambda_0\lambda_1+\dots \ .
}

The action is Poincare invariant, and hence only depends on the Mandelstam invariants $s=-(k_1+k_2)^2$ and $t=-(k_1+k_4)^2$. We can always do a Poincare transformation that places the four external momenta in an $R^{1,2}$ subspace of $R^{1,d-1}$. A convenient choice is
\eqn\momenta{\eqalign{
k_1&=\left({\sqrt s\over2},\ \ {\sqrt s\over2},0\right)\ ,\quad  k_3=\left(-{\sqrt s\over2},\ \ \,{s+2t\over2\sqrt{s}},\ \ i{\sqrt{t(s+t)}\over\sqrt s}\right)\cr
k_2&=\left({\sqrt s\over2},-{\sqrt s\over2},0\right)\ ,\quad k_4=\left(-{\sqrt s\over2},-{s+2t\over2\sqrt{s}},-i{\sqrt{t(s+t)}\over\sqrt s}\right)
}}
where the signature is $(-,+,+)$. Note that the third component is purely imaginary. This is because we are considering imaginary scattering angles, where $s,t>0$. As mentioned in the main text, one may think of the momenta \momenta\ as being real, with signature $(-,+,-)$.

\subsec{Order $m^0$}

At leading order $m=0$ and the action becomes free. The corresponding classical solution in this case is the famous one of \refs{\GrossKZA,\GrossGE}, (or its analog for $s,t>0$). Here, we present it as a preparation for higher orders.

At zero mass the equations of motion simplify to
\eqn\zerotheom{
\d\bar\d x_0^\mu=0\ ,\qquad{1\over2\pi\alpha'}\left.\d_\tau x_0^\mu\right|_{\tau=0}=i\sum_jk_j^\mu\delta(\sigma-\sigma_j)\ ,\qquad e_0=0 \ .
}
Correspondingly, the free classical solution is given by
\eqn\GMsolution{
x_0^\mu=i\alpha'\sum_jk^\mu_j\log|z-\sigma_j|^2 \ .
}
By plugging this back into the free action we arrive at
\eqn\zeroorder{
S_0={1\over4\pi\alpha'}\int d\sigma d\tau\,\d_\alpha x_0\cdot\d_\alpha x_0+i\sum_jk_j\cdot x_0(\sigma_j)=-{1\over2}\alpha'\left[t\log(1+\lambda_0)+s\log(1+1/\lambda_0)\right]
}
Finally, extremizing $S_0$ with respect to $\lambda_0$ we find that $\lambda_0=s/t$. By plugging this back into the action we get
\eqn\actionzeroodred{
S_0= \alpha' \left[s \log s+t \log t-(s+t) \log (s+t) \right] 
}
in agreement with \limitVen.

Note that we did not have to impose the Virasoro constraints
\eqn\virasorozerothorder{
\d_z x_0\cdot\d_zx_0=\d_{\bar z}x_0\cdot\d_{\bar z}x_0=0 \ .
}
It is automatically satisfied after we minimize the action \zeroorder\ with respect to the vertices insertion points. This will be true for higher orders in $m$ as well.

\subsec{Order $m^{1/2}$}
This order simply vanishes by the e.o.m. for $x_0$ and $\lambda_0$. Explicitly, we have
\eqn\firstorder{
S_1={2\over4\pi\alpha'}\int d\sigma d\tau\,\d_\alpha x_0\cdot\d_\alpha x_1+i\sum_jk_j\cdot x_1(\sigma_j)+i\lambda_1\d_\lambda \sum_jk_j\cdot \left.x_0(\check\sigma_j)\right|_{\lambda=0}
}
where $\check\sigma$ was defined in \lambdagauge.

The first two terms are zero because $x_0^{\mu}$ is on-shell. The last term is zero because $S_0$ was minimized with respect to variations of $\lambda_0$. Hence, $S_1=0$ and at this order we get no constraint on $x_1^{\mu}$, $\lambda_1$.

\subsec{Order $m^{1}$}

At order $m^1$ we have
\eqn\secondorder{\eqalign{
S_2&={2\over4\pi\alpha'}\int d\sigma d\tau\,\d_\alpha x_0\cdot\d_\alpha x_2+i\sum_jk_j\cdot x_2(\sigma_j)+i\lambda_2\d_\lambda \sum_jk_j\cdot \left.x_0(\check\sigma_j)\right|_{\lambda=0}\cr
&+{1\over4\pi\alpha'}\int d\sigma d\tau\,\d_\alpha x_1\cdot\d_\alpha x_1+\int d\sigma\, e_1\,\d_\sigma x_0\cdot\d_\sigma x_1+i\lambda_1\d_\lambda \sum_jk_j\cdot \left.x_1(\check\sigma_j)\right|_{\lambda=0}\cr
&+{i\over2}\lambda_1^2\d_\lambda^2 \sum_jk_j\cdot \left.x_0(\check\sigma_j)\right|_{\lambda=0} \ .
}}
The first line vanishes for the same reason as \firstorder . Indeed,  exchanging $x_1^{\mu}\to x_2^{\mu}$ and $\lambda_1\to\lambda_2$ the two are identical. In the second line we have a free bulk field, $x_1^{\mu}$ with two boundary sources. We split it accordingly 
\eqn\xsplit{
x_1^{\mu}=\tilde x_1^{\mu}+y_1^{\mu} \ ,
}
where
\eqn\xone{
y_1^\mu(z,\bar z)=-\alpha'\int d\sigma\,\d_\sigma\left(e\,\d_\sigma x_0^\mu\right)\log|z-\sigma|^2=\alpha'\int d\sigma\, e_1\,\d_\sigma x_0^\mu\left({1\over\sigma-z}+{1\over\sigma-\bar z}\right) \ ,
}
and
\eqn\yone{
\tilde x_1^{\mu} (z,\bar z)=i{\alpha'\over2}k_1^{\mu}\,\lambda_1\left({1\over z-\sigma_1}+{1\over\bar z-\sigma_1}\right) .
}
We see that $\tilde x_1^{\mu}\propto k_1^{\mu}$ and therefore $\sum k_j\cdot \tilde x_1(\check\sigma_j)$ is independent of $\lambda$. As a result
\eqn\somezeroterms{
{1\over2\pi\alpha'}\int d^2z\d\tilde x_1\cdot\bar\d\tilde x_1+i\lambda_1\d_\lambda \sum_jk_j\cdot\left.\tilde x_1(\check\sigma_i)\right|_{\lambda=0}=0 \ .
}
Next, we have 
\eqn\morezeros{
{2\over2\pi\alpha'}\int d^2z\d\tilde x_1\cdot\bar\d\tilde y_1+\int d\sigma\, e_1\,\d_\sigma x_0\cdot\d_\sigma\tilde x_1=0 \ ,
}
where we have viewed $\tilde x_1^{\mu}$ as a perturbation of $y_1^{\mu}$. We remain with three terms
\eqn\threeterms{
S_2=i\lambda_1\d_\lambda\sum_jk_j\cdot y_1(\sigma_j)+{i\over2}\lambda_1^2\d_\lambda^2 \sum_jk_j\cdot \left.x_0(\check\sigma_j)\right|_{\lambda=0}+{1\over2}\int d\sigma\, e_1\,\d_\sigma x_0\cdot\d_\sigma y_1 \ .
}
They evaluate to
\eqn\threetermsare{\eqalign{
{i\over2}\lambda_1^2\d_\lambda^2 \sum_jk_j\cdot \left.x_0(\check\sigma_i)\right|_{\lambda=0}&=-{\alpha'\over2}{s\, t\over s+t}\lambda_1^2 , \cr
i\lambda_1\d_\lambda\sum_jk_j\cdot y_1(\sigma_j)&=-2\alpha'\sqrt{s t}\,E[e_1]\lambda_1 , \cr
{1\over2}\int d\sigma\, e_1\,\d_\sigma x_0\cdot\d_\sigma y_1&=-2\alpha'(s+t)E[e_1]^2 ,
}}
where
\eqn\Edefinition{
E[e]\equiv\int d\sigma\, \mu(\sigma)\ ,\qquad\mu(\sigma) \equiv \alpha'{\sqrt{\prod|\sigma_{i+1}-\sigma_i|}\over\prod\left(\sigma-\sigma_j\right)}e_1(\sigma) . 
}
Combining all pieces together we get that $S_2$ is a complete square
\eqn\Stwoonshell{
S_2=-\alpha'{s\,t\over2(s+t)}\left(\lambda_1+{2(s+t)\over\sqrt{s\,t}}E[e_1]\right)^2 .
}
Extremizing \Stwoonshell\ with respect to $\lambda_1$, we arrive at
\eqn\solone{
\lambda_1=-2{s+t\over\sqrt{s\,t}}E[e_1] .
}
We conclude that
\eqn\Stwo{
S_2=0 \ ,
}
and, hence, the correction starts at order $m^{3/2}$.

Note that \solone\ is precisely the Virasoro constraint at order $\sqrt{m}$. Namely,
\eqn\Virtwo{
\d_z x_1\cdot\d_z x_0=-{\alpha'\over2}{\sqrt{\prod|\sigma_{i+1}-\sigma_i|}\over\prod\left(z-\sigma_j\right)}\left(\sqrt{s\,t}\,\lambda_1+2(s+t)E[e_1]\right)
}
is automatically satisfied once we extremize with respect to all variables.

\subsec{Order $m^{3/2}$}

At order $m^{3/2}$ we organize all the contributions as follows
\eqn\Sthree{
S_3={1\over2}\int d\sigma{m^2\over e_1}+A_1+A_2+A_3+A_4
}
where
\eqn\Aterms{\eqalign{
A_1&={1\over2\pi\alpha'}\int d\sigma d\tau\,\d_\alpha x_0\cdot\d_\alpha x_3+i\sum_jk_j\cdot\left(x_3(\sigma_j)+i \lambda_3\d_\lambda \left.x_0(\check\sigma_j)\right|_{\lambda=0}\right) , \cr
A_2&={1\over2\pi\alpha'}\int d\sigma d\tau\,\d_\alpha x_1\cdot\d_\alpha x_2+\int d\sigma\, e_1\,\d_\sigma x_0\cdot\d_\sigma x_2+i\lambda_1\d_\lambda\sum_jk_j\cdot  \left. x_2(\check\sigma_j)\right|_{\lambda=0} , \cr
A_3&=\int d\sigma\, e_2\,\d_\sigma x_1\cdot\d_\sigma x_0+i \lambda_2 \d_\lambda \sum_jk_j\cdot \left.x_1(\check\sigma_j)\right|_{\lambda=0}+i \lambda_2 \lambda_1\d_\lambda^2 \sum_jk_j\cdot \left.x_0(\check\sigma_j)\right|_{\lambda=0} , \cr
A_4&={1\over2}\int d\sigma\, e_1\,\d_\sigma x_1\cdot\d_\sigma x_1+{i\over2!}\lambda_1^2\d_\lambda^2 \sum_jk_j\cdot\left.x_1(\check\sigma_j)\right|_{\lambda=0}+{i\over3!}\lambda_1^3\d_\lambda^3 \sum_jk_j\cdot \left.x_0(\check\sigma_j)\right|_{\lambda=0} \ .
}}

We find that:
\item{1)} $A_1=0$ because $x_0$ and $\lambda_0$ are the extremum of $S_0$. $A_1$ is a linear variation of $S_0$ around this extremum. 
\item{2)} $A_2=0$ for any $x_2$. This is because $x_1$ is the extremum of $S_2$, \secondorder, and in $A_2$ we have a linear variation of $x_1$ in $S_2$ around this point.
\item{3)} The first term in $A_3$ vanishes because $\d_\sigma x_1\cdot\d_\sigma x_0=0$. The last two terms sum up to zero because
\eqn\sumtozero{
i \lambda_2 \d_\lambda \sum_jk_j\cdot \left.x_1(\check\sigma_j)\right|_{\lambda=0}=-i \lambda_2\lambda_1\d_\lambda^2 \sum_jk_j\cdot \left.x_0(\check\sigma_j)\right|_{\lambda=0}={\alpha'\over2} \lambda_2 \sqrt{s\,t}\,E[e_1]
}
The reason for this cancellation is that at order $m^1$ we concluded that $S_2$ is given by
\eqn\secondorderonshell{\eqalign{
S_2&={1\over4\pi\alpha'}\int d\sigma d\tau\,\d_\alpha x_1\cdot\d_\alpha x_1+\int d\sigma\, e_1\,\d_\sigma x_0\cdot\d_\sigma x_1\cr
&+i\lambda_1\d_\lambda \sum_jk_j\cdot  \left.x_1(\check\sigma_i)\right|_{\lambda=0}
+{1\over2}\lambda_1^2\d_\lambda^2\sum_jk_j\cdot  \left.x_0(\check\sigma_i)\right|_{\lambda=0}\cr
&=-\alpha'{s\,t\over2(s+t)}\left(\lambda_1+{2(s+t)\over\sqrt{s\,t}}E[e_1]\right)^2 \ .
}}
If we vary this as $\lambda_1\to\lambda_1+\lambda_2$ or $e_1\to e_1+e_2$ to linear order and adjust the classical solution accordingly $x_1^{\mu}\to x_1^{\mu}+\delta x^{\mu}(e_1,\lambda_2)$ then the result vanishes because we started at the extremum. As we seen above, any linear variation $x_1^{\mu}\to x_1^{\mu}+\delta x^{\mu}$ of $S_2$ vanishes by itself, just the same as $A_2=0$. So we see that $A_3=0$ too. Note that $S_3$ is therefore independent of $e_2$ and $\lambda_2$. Hence, we can evaluate it without solving for these new variables.

\item{4)} The three terms in $A_4$ are given by
\eqn\Afourterms{\eqalign{
{i\over3!}\lambda_1^3\d_\lambda^3 \sum_jk_j\cdot  \left.x_0(\check\sigma_i)\right|_{\lambda=0}&=-{\alpha'\over6}{s+t\over\sqrt{s\,t}}(5s+4t)E[e_1]^3 \ , \cr
{i\over2!}\lambda_1^2\d_\lambda^2 \sum_jk_j\cdot  \left.x_1(\check\sigma_i)\right|_{\lambda=0}&={\alpha'\over2}{(s+t)^2\over\sqrt{s\,t}}E[e_1]^2\int d\sigma{\mu(\sigma)\over\sigma-\sigma_1}\ ,\cr
{1\over2}\int d\sigma\, e_1\,\d_\sigma x_1\cdot\d_\sigma x_1&=-{\alpha'^2\over4}(s+t)\int\!d\sigma\,\mu(\sigma)\int\!d\chi\,\mu(\chi)\cr
&\times\int\! d\xi\, e_1(\xi)\left({(\sigma-\chi)^2\over(\sigma-\xi)^2(\xi-\chi)^2}-{2(\sigma-\sigma_1)^2\over(\sigma-\xi)^2(\xi-\sigma_1)^2}\right) \ .
}}

Note that the on-shell action is precisely of the form \cubic\ that we used in the text. In this way, we can run a simple argument and conclude that the result is given by \onshellaction. 

To run the argument, however, we should make sure that \Afourterms\ is not zero. Surprisingly, if we symmetrize the sum of three terms above over integration variables the sum vanishes for any $e_1$.\foot{We thank T. Cohen for pointing this to us.} The integrals, however, contain divergences on the integration contour and we need to treat them carefully with $i\epsilon$ prescription coming from the worldsheet. After this is done the result is indeed what we expected. Showing this explicitly is the subject of the next section.

\subsec{Evaluation of $A_4$}

We now evaluate $A_4$ on the solution for $e_1(\sigma)$ \efourpoint\ explicitly and confirm the result \mainagain. All we need for the derivation in the main text is that $A_4\neq 0$. Still, it does not harm to confirm the result by an independent explicit calculation.

To evaluate the integrals in \Afourterms\ explicitly it is convenient to introduce the following integral
\eqn\integral{
{\cal I}(z)= \int{d \sigma\over \sigma - z}   {\ \prod_i |\sigma - \sigma_i|^{1/2} \over \prod_i (\sigma - \sigma_i )} = \sum_{i=1}^{4} (-1)^i {\cal I}_i (z)\ ,\qquad
{\cal I}_i (z) = \int\limits_{\sigma_i}^{\sigma_{i+1}}{d \sigma\over \sigma - z} {1 \over \prod_j |\sigma - \sigma_j|^{1/2}}  \ ,
}
where we assume $z$ to be in the upper half-plane. It is also convenient to introduce
\eqn\definition{
E \equiv -\lim_{z \to \infty} z\,{\cal I}(z)= \int d \sigma  {\ \prod_i |\sigma - \sigma_i|^{1/2} \over \prod_i (\sigma - \sigma_i )}=4 \sqrt{{2(2 s + t) \over s +t} } \left[ K \left({t \over s+t} \right) - K \left({s \over s + t} \right) \right] \ .
}

To perform the integral in ${\cal I}_i (z)$, \integral, we make the following change of variables
\eqn\changeofv{
\sigma = \sigma_i + (\sigma_{i+1} - \sigma_{i}) \sin^2{\phi}\quad\Rightarrow\quad {d \sigma \over \sqrt{\sigma- \sigma_i} \sqrt{\sigma_{i+1} - \sigma}} = 2 d \phi
}
where $0 \leq \phi \leq {\pi \over 2}$. After this change ${\cal I}_i (z)$ becomes
\eqn\resultB{\eqalign{
{\cal I}_i (z) &=- {2 \over (\sigma_{i+1} - \sigma_i)^2} \int\limits_0^{{\pi \over 2}}\! d \phi\, {1 \over \sqrt{{\sigma_{i+1} - \sigma_{i - 1}\over \sigma_{i+1} - \sigma_{i}} - \cos^2 \phi}}  {1 \over \sqrt{{\sigma_{i+2} - \sigma_{i +1}\over \sigma_{i+1} - \sigma_{i}} + \cos^2 \phi}} {1 \over {z - \sigma_{i+1} \over \sigma_{i+1} - \sigma_{i}} +  \cos^2 \phi}\cr
&={2 \over |\sigma_{i+1} - \sigma_{i-1}| |\sigma_{i+2} - \sigma_i| (\sigma_{i+2} - z) } \left( K(u)+ {\sigma_{i+2} - \sigma_i \over \sigma_i - z}  {1 \over \sqrt{1- u}} \Pi(-w , {u \over u - 1}) \right)
}}
where
\eqn\uzdefinition{
u={(\sigma_{i+2} - \sigma_{i+1})(\sigma_{i} - \sigma_{i-1}) \over (\sigma_{i+2} - \sigma_{i})(\sigma_{i+1} - \sigma_{i-1})}\ ,\qquad w ={(z - \sigma_{i+2}) (\sigma_{i+1} - \sigma_i) \over (z - \sigma_{i}) (\sigma_{i+2} - \sigma_{i+1})} .
}

We can now plug the solution for $e_1(\sigma)$ \efourpoint\ into \Afourterms\ to get

\eqn\relationto{\eqalign{
&-{\alpha'^2\over4}(s+t)\int\!d\sigma\,\mu(\sigma)\int\!d\chi\,\mu(\chi) \int\! d\xi\, e_1(\xi) {(\sigma-\chi)^2\over(\sigma-\xi)^2(\xi-\chi)^2} \cr
&= \alpha'^2 {\sqrt{s t (s+t)} \over \pi (2 s +t)} \int d \xi \ e_1(\xi) \  \left( {\rm Re}[{\cal I}(\xi)]^2 - E \ {\rm Re}\left[{d {\cal I}(\xi)\over d \xi } \right]  \right), \cr
\cr
&-{\alpha'^2\over4}(s+t)\int\!d\sigma\,\mu(\sigma)\int\!d\chi\,\mu(\chi) \int\! d\xi\, e_1(\xi) \left( -{2(\sigma-\sigma_1)^2\over(\sigma-\xi)^2(\xi-\sigma_1)^2} \right) \cr
&= \alpha'^2 {\sqrt{s t (s+t)} \over \pi (2 s +t)}  \int d \xi \ e_1(\xi) \  E \ \left( {\rm Re}\left[{d {\cal I}(\xi)\over d \xi } \right]   +  {2 {\rm Re}[{\cal I}(\xi)] \over \sigma_1 - \xi} + {E \over (\sigma_1 - \xi)^2} \right) \ , \cr
\cr
&{\alpha'\over2}{(s+t)^2\over\sqrt{s\,t}}E[e_1]^2\int d\xi{\mu(\xi)\over\xi-\sigma_1} =- \alpha'^2 {\sqrt{s t} (s+t)^{3/2} \over 2 \pi (2 s +t)^2 } \int d\xi \ {e_1(\xi) \over \prod (\xi - \sigma_i)} {1 \over \sigma_1 - \xi} ,
}}
where to get these results one should be careful with the $i \eps$ prescription which we kept implicit in the formulas for the on-shell action. We perform the last integral over $\xi$ numerically. To do this we need to understand the behavior of the function close to the vertex operators. The result is that it has a $(\sigma - \sigma_i)^{- {1 \over 2}}$ integrable singularity close to $\sigma_{2,3,4}$ and has a simple pole at $\sigma_1$. These statements are not manifest when looking at \relationto . For example the last line has a singularity ${1 \over (\xi - \sigma_1)^{3/2}}$ close to $\xi = \sigma_1$, it however is canceled against the other terms. The $i \eps$ prescription instructs us to treat the pole at $\xi = \sigma_1$ as a principal value. When combined with the first line in \Afourterms\ we get that
\eqn\result{
A_4 = {1 \over 6} \int d \sigma {m^2 \over e_1} \qquad\Rightarrow\qquad
S_3 ={1 \over 2} \int d \sigma {m^2 \over e_1} + A_4 = {2 \over 3} \int d \sigma {m^2 \over e_1}  \ ,
}
as expected.

\newsec{Appendix B: Bootstrap in Mellin Space}

It is natural to ask what is the analog of the flat space S-matrix asymptotic bootstrap \CaronHuotICG\ in AdS. A natural idea is to consider the Mellin amplitude $M(s,t)$ and focus on the regime where $s$ and $t$ are larger than any relevant scale in the problem, say $\sqrt \lambda$. Let us recall what is the Mellin representation of the planar connected four-point function $F(u,v)$ \MackMI , \PenedonesUE. The standard version of Mellin transform takes the form
\eqn\mellin{\eqalign{
\widehat M(\gamma_{12}, \gamma_{14}) &= \int\limits_0^{\infty} {d u d v \over u v} \ u^{\gamma_{12}} v^{\gamma_{14}} F(u,v)\ .
}}
We imagine that this integral converges for $a < {\rm Re} [\gamma_{12}, \gamma_{14} ] < b$. Then we can write the inverse transformation
\eqn\inversemellin{
F(u,v) = \int\limits_{c - i \infty}^{c + i \infty} {d \gamma_{12} d \gamma_{14} \over (2 \pi i)^2} \ u^{- \gamma_{12}} v^{- \gamma_{14} } \widehat M (\gamma_{12}, \gamma_{14})\ ,\qquad a<c<b\ .
}
The relation between $\gamma_{ij}$ and Mandelstam-like variables we mentioned above is the following \CostaCB
\eqn\relationvar{
\gamma_{12} =\Delta - {t \over 2}\ ,\qquad \gamma_{13} =- {s \over 2} \ ,\qquad \gamma_{14} = \Delta - \gamma_{12} - \gamma_{13} = {s + t \over 2} =  - {u \over 2} \ ,
}
where $\Delta$ is the conformal dimension of the external identical scalar operators.

In the discussion of planar theories it is common to define the Mellin amplitude by writing down explicitly the gamma-functions pre-factor
\eqn\largeNmellin{
\widehat M(\gamma_{12}, \gamma_{14}) = \left[ \Gamma(\gamma_{12}) \Gamma(\gamma_{13}) \Gamma(\gamma_{14}) \right]^2 M(s,t)\ .
} 
The pre-factor has singularities at $\gamma_{12}, \gamma_{13}, \gamma_{14} = - {\rm integer}$, which corresponds to the exchange of the double trace operators.

As is the case with the usual scattering amplitudes, the Mellin amplitude $M(s,t)$ has poles dictated by the twists of the exchanged operators $\tau = \Delta - J$. The residues are fixed in terms of $\Delta$ and $J$ and the correspondent three-point couplings. It takes the form
\eqn\poles{
M(s,t) \simeq {c_{\Delta \Delta \tau}^2 {\cal Q}_{J,m}^{\Delta,\tau, d}(s)\over t - (\tau + 2 m)} +\dots\ , \qquad  {\cal Q}^{\Delta,\tau, d}_{J,m}(s) = K(\Delta, J, m) \ Q^{\Delta,\tau, d}_{J,m}(s)\ ,
}
where $m = 0,1,2,\dots$ stands for the level of the descendent, $Q^{\Delta,\tau, d}_{J,m}(s) $ are polynomials in $s$ of degree $J$, $c_{\Delta \Delta \tau}^2$ labels the correspondent three-point coupling and $K(\Delta, J, m)$ is a kinematical pre-factor that can be found in \CostaCB.  As in flat space, unitarity implies that for $s,t \gg 1$ the Mellin amplitude develops large imaginary part. 

Let us assume that the states that dominate the Mellin amplitude have large spin $J \gg s,t$. In this case the crossing equation to leading order takes the form
\eqn\crossingleading{
M(s,t) \simeq \sum_j^{J(t)} c_j\, j^s = \sum_j^{J(s)} c_j\, j^t \simeq M(t,s)\ , 
}
where we have used the large $J$ asymptotic of the Mack Polynomials
\eqn\fullresidue{
\lim_{j\to\infty}{\cal Q}^{\Delta,\tau, d}_{J,m}(s) = - {2^{J-m+\tau} \over \sqrt \pi \ \Gamma(m+1) \Gamma({2 \Delta - 2 m - \tau \over 2})^2 } {J^{s+{1 \over 2}} ( s^m+ O(s^{m-1}) ) \over \Gamma({s + \tau + 2 m \over 2})^2} + O({1 \over J})  \ .
}

The crossing equation \crossingleading\ is basically telling us that $J(t)^s\simeq J(s)^t$. It has the trivial solution
\eqn\solutionM{
J(t) = e^{ {t \over c} }\ . 
}
A posteriori we see that, indeed, $J(t) \gg t$ as we assumed.

Notice that close to the pole at $t = \tau + 2m$, the Mellin amplitude is controlled by $J(t) = J(\tau + 2m)$. At large twist, the solution \solutionM\ corresponds to
\eqn\twistcrossing{
\tau(J) =  c \log J +\dots\ . 
}
Of course, this formula is very familiar from the studies of the gauge theories \AldayMF. Here we see how this result comes out of the simple bootstrap consideration. It implies that the asymptotic behavior of the Mellin amplitude takes the form
\eqn\asympt{
\lim_{s,t\to\infty}\log M(s,t) = {1 \over c} \,s\, t +\dots\ ,
}
Identical formulas can be written for the other channels: $s,u \gg 1$ or $t,u \gg 1$. 

This result is related to the double light-cone limit of the connected four-point function discussed in \AldayZY. Indeed, the limit $\gamma_{12}, \gamma_{14}\to\infty$ of \mellin\ is controlled by the $u,v \to 0$ of $F(u,v)$. In \AldayZY\ it was argued that in planar gauge theory, this limit is controlled by the factor
\eqn\AMlimit{
\lim_{u,v\to0}F(u,v)\simeq e^{- {\Gamma_{cusp} \over 4} \log u \log v }\ ,
}
which is related to the Alday-Maldacena solution \AldayHR. Let us see what this behavior implies in Mellin space. 

We start from the kinematical region $0 \leq u,v \leq 1$ consider the small $u,v$ contribution to $\widehat M(\gamma_{12}, \gamma_{14})$ in \mellin. Using \AMlimit\ we arrive at \eqn\mellincusp{
\widehat M(\gamma_{12}, \gamma_{14}) \simeq \int\limits_0^1 {d u\, d v \over u\, v} u^{\gamma_{12}} v^{\gamma_{14}} e^{- {\Gamma_{cusp} \over 4} \log u \log v } = \int\limits_0^{\infty} d V {e^{-V} \over \gamma_{12} \gamma_{14} + {\Gamma_{cusp} \over 4} V }  \  .
}
This integral is well-defined in the $\gamma_{12}, \gamma_{14}>0$ region. To connect with \asympt\ we analytically continue to the region $\gamma_{12}, \gamma_{14} < 0$ (and hence positive $s$ and $t$). As we analytically continue $\gamma_{12}$ and $\gamma_{14}$, the pole in \mellincusp\ moves from the negative $V$-axis towards the positive axis and then back to the negative axis dragging the contour of integration together with it. The leading asymptotic is given by the residue at the pole. It gives
\eqn\leadingasymptotic{
\lim_{s,t \to  \infty}\log M(s,t) = {1 \over \Gamma_{cusp} } s\, t +\dots\ .
}
Comparing this with \asympt\ we learn that $c = \Gamma_{cusp}$. This is precisely what we expect from \twistcrossing. Notice that regions other than $0 \leq u, v \leq 1$ produce smaller contributions at large $\gamma_{12}, \gamma_{14} < 0$, which justifies our approximation above.

Let us conclude this appendix with two future directions. First, it would be interesting to understand how the vector models and SYK-like models fit into this picture. In these cases, it is not true that $J(t) \gg t$ and there is an accumulation point in the twist spectrum. Finally, it would also be interesting to understand the relation of our considerations in this appendix to the recent work on bootstrap in Mellin space \refs{\GopakumarWKT,\GopakumarCPB}.

\listrefs

\bye